\documentclass[fleqn,usenatbib]{mnras}

\usepackage{newtxtext,newtxmath}

\usepackage[T1]{fontenc}

\DeclareRobustCommand{\VAN}[3]{#2}
\let\VANthebibliography\thebibliography
\def\thebibliography{\DeclareRobustCommand{\VAN}[3]{##3}\VANthebibliography}

\usepackage{graphicx}	
\usepackage{amsmath}	
\usepackage{xspace}

\usepackage{comment}

\newcommand{\msun}{{\,\rm M_\odot}}

\newcommand{\kms}{\,{\rm km}\,{\rm s}^{-1}}

\newcommand{\Myr}{\,{\rm Myr}}

\newcommand{\Mpc}{\,{\rm Mpc}}

\newcommand{\mmag}{\,{\rm mag}}

\newcommand{\thesanzoom}{\textsc{thesan-zoom}\xspace}

\def\jcap{J. Cosmol.  Astropart. Phys.}
\def\aap{A\&A}
\def\apj{ApJ}

\def\apjl{ApJ}
\def\mnras{MNRAS}
\def\araa{ARA\&A}
\def\aj{AJ}

\def\physrep{Phys. Rep.}
\def\nat{Nature}

\def\apjs{ApJS}
\def\prd{Phys. Rev. D}

\title[Rapid Early Structure Formation]{The Cosmic Rush Hour: Rapid Formation of Bright, Massive, Disky, Star-Forming Galaxies as Signatures of Early-Universe Physics}

\author[Shen et al.]{\parbox{17.5cm}{
Xuejian Shen,$^{1}$\thanks{E-mail: xuejian@mit.edu}
Oliver Zier,$^{2,1}$
Mark Vogelsberger,$^{1}$
Michael Boylan-Kolchin,$^{3}$
Lars Hernquist,$^{2}$
Sandro Tacchella,$^{4,5}$
Rohan P. Naidu$^{1}$\thanks{NASA Hubble Fellow}
}
\\ \vspace{0.2cm} \\
$^{1}$ Department of Physics, Kavli Institute for Astrophysics and Space Research, Massachusetts Institute of Technology, Cambridge, MA 02139, USA \\
$^{2}$ Center for Astrophysics | Harvard \& Smithsonian, 60 Garden St, Cambridge, MA 02138, USA \\
$^{3}$ Cosmic Frontier Center, Department of Astronomy, The University of Texas at Austin, 2515 Speedway Stop C1400, Austin, TX 78712, USA \\ 
$^{4}$ Kavli Institute for Cosmology, University of Cambridge, Madingley Road, Cambridge, CB3 0HA, UK \\
$^{5}$ Cavendish Laboratory, University of Cambridge, 19 JJ Thomson Avenue, Cambridge, CB3 0HE, UK \\
}

\date{Accepted XXX. Received YYY; in original form ZZZ}

\pubyear{\the\year{}}

\begin{document}
\label{firstpage}
\pagerange{\pageref{firstpage}--\pageref{lastpage}}
\maketitle

\begin{abstract}
Early JWST observations have revealed a high-redshift universe more vibrant than predicted by canonical galaxy-formation models within $\Lambda$CDM, showing an excess of ultraviolet(UV)-bright, massive, and morphologically mature galaxies. Departures from $\Lambda$CDM prior to recombination can imprint signatures on non-linear structure formation at high redshift. In this paper, we investigate one such scenario -- Early Dark Energy, originally proposed to resolve the Hubble tension -- and its implications for these high-redshift challenges. We present the first large-scale cosmological hydrodynamic simulations of these models. Modifications to the pre-recombination expansion history accelerate early structure formation and produce UV luminosity and stellar mass functions in excellent agreement with JWST measurements, requiring essentially no additional calibrations. Predictions converge to $\Lambda$CDM at lower redshifts ($z \lesssim 3$), thereby preserving all successes of $\Lambda$CDM. This model also accelerates the emergence of stellar and gaseous disks, increasing their number densities by $\sim 0.5$ dex at $z\simeq 6$--7, primarily due to the higher abundance of massive galaxies. Taken together, these results demonstrate how early-universe physics can simultaneously reconcile multiple high-redshift challenges and the Hubble tension while retaining the core achievements of $\Lambda$CDM. This opens a pathway to constraining a broad class of beyond-$\Lambda$CDM models with forthcoming observations. 
\end{abstract}




\section{Introduction}

The first few years of observations by the James Webb Space Telescope (JWST) have revealed a significantly more vigorous early universe than anticipated from extrapolations of lower-redshift studies. Many of these findings challenge the canonical picture of galaxy formation and evolution and may point toward physics beyond the standard $\Lambda$CDM cosmological model. One of the most striking signatures of more rapid structure formation is the elevated abundance of ultraviolet (UV)-bright galaxies at cosmic dawn ($z \gtrsim 10$), a trend that was evident as early as the initial JWST data release~\citep[e.g.][]{Naidu2022,Castellano2022,Finkelstein2022,Adams2023,Atek2023,Bouwens2023a,Donnan2023,Harikane2023,Robertson2023,Yan2023,Hainline2024} and later confirmed through spectroscopic observations extending out to $z \sim 14$~\citep[e.g.][]{Curtis2023,Carniani2024,Roberts-Borsani2024,Harikane2024-spec,Harikane2024b-spec}. The large number densities of these bright galaxies are in tension with predictions of the majority of the pre-JWST theoretical models~\citep[as summarized in e.g.][]{Finkelstein2023,Finkelstein2024}. These unexpected results have sparked widespread discussions of potential explanations, including Eddington bias amplified by bursty star-formation~\citep[e.g.][]{Mason2023,Mirocha2023,Shen2023,Sun2023}, top-heavy stellar initial mass functions~\citep[IMF; e.g.][]{Inayoshi2022,Yung2023,Cueto2023,Trinca2024,Lu2024}, systematically enhanced star-formation efficiency in extreme environments~\citep[e.g.][]{Dekel2023,Li2023,BK2025,Shen2025}, and modifications to early-universe cosmology~\citep[e.g.][]{Para2023,Sabti2024,Shen2024b}, among others. However, substantial degeneracies persist among these solutions, underscoring the need for additional observables to constrain the galaxy–halo connection at high redshifts~\citep[e.g.][]{Munoz2023,Shen2024b,Gelli2024}.

This ``UV luminosity function'' tension is accompanied by the discovery of several extremely massive galaxy candidates across a broader redshift range ($z \sim 3-12$; e.g. \citealt{Labbe2023,Akins2023,Xiao2023,deGraaff2024,Casey2024,Glazebrook2024}), raising concerns about whether such systems could have assembled their stellar mass given the available baryonic reservoir in a $\Lambda$CDM universe (\citealt{BK2023,Lovell2023}, though see \citealt{Cochrane2025}). Many of these sources were later either identified as low-redshift interlopers or showed potentially large contributions from active galactic nuclei (AGN), with stellar mass estimates further complicated by uncertainties in the assumed IMF, star-formation histories, and the contribution of nebular emission~\citep[e.g.][]{Endsley2022,Larson2022,Kocevski2023,Desprez2024,Narayanan2024,Wang2024a,Wang2024c,Turner2025}. Subsequent constraints combining large-area JWST survey programs have loosened the tension in terms of individual very massive galaxies. Nevertheless, these studies still found number densities of massive galaxies exceeding the predictions of pre-JWST theoretical models~\citep[e.g.][]{Weibel2024, WangT2024, Harvey2025}, which require at least moderately enhanced star-formation efficiency at high redshift or tuning of the cosmological model. Notably, a non-negligible population of heavily dust-obscured massive galaxies, missed by traditional Lyman-break selection using the Hubble Space Telescope, contributes to the massive end.

Besides the abundance of massive and bright galaxies, JWST has revealed the unexpectedly early emergence of morphologically and dynamically mature systems at high redshifts. Here, ``mature'' refers to a sequence well-established from low-redshift studies: galaxies evolve from low-mass, irregular, dispersion-supported dwarfs to more massive, rotationally supported disky galaxies~\citep[e.g.][]{El-Badry2018,Tacchella2019,Thob2019,Gurvich2023}. This progression is supported by decades of galaxy surveys~\citep[e.g.][]{Emsellem2007,Obreschkow2014,Wisnioski2015,Simons2015,Cortese2016,Lange2016,Clauwens2018,Tiley2021} and by the chemo-kinematic properties of stars in the Milky Way~\citep[e.g.][]{Johnson2021,Belokurov2022,Conroy2022,Semenov2024,
Chandra2024}. Early JWST imaging data indicate that a substantial fraction of high-redshift galaxies exhibit edge-on, disk-like morphologies with low S\'{e}rsic indices~\citep[e.g.][]{Robertson2023b, Ferreira2023, Kartaltepe2023, Huertas-Company2024, SunW2024} and bars~\citep[e.g.][]{Bland-Hawthorn2023-bar,Guo2025}. For example, \citet{Kartaltepe2023} reported that the fraction of ``disky'' galaxies maintains nearly $30\%$ out to $z \sim 6-9$. Complementary ALMA observations of bright, lensed galaxies have also detected dynamically cold, rotating gaseous disks out to $z \gtrsim 7$~\citep[e.g.][]{Rizzo2021,Jones2021,Roman-Oliveira2023,Fujimoto2024,Rowland2024}, characterized by high rotation-to-dispersion ratios ($V_{\rm rot}/\sigma \gtrsim 10$) and prominent clumpy substructures~\citep[e.g.][]{Fujimoto2024}. Similar findings have been reported using ionized gas probed by JWST as well~\citep[e.g.][]{Nelson2024,Xu2024,Danhaive2025}. 

Another type of ``mature'' system is quenched galaxies, which appear as galaxies exceed substantially the Milky Way mass~\citep[e.g.][]{Bell2003,Kauffmann2003,Faber2007,Peng2010,Muzzin2013,Moster2018,Behroozi2019}.
JWST has uncovered a surprisingly large population of quenched galaxies already in place as early as $z \simeq 7$~\citep[e.g.][]{Carnall2023a,Carnall2023b,Valentino2023,Weaver2023,Alberts2024,DeGraaff2025,Weibel2025}, implying that the mechanisms responsible for suppressing star-formation were active much earlier than previously anticipated~\citep[e.g.][]{Chittenden2025,Lagos2025}. A coherent explanation for the subtle timing of the emergence of these ``mature'' systems and the more apparent galaxy abundance discrepancies remains to be identified.

Meanwhile, from the theory side, a range of early-universe physics that perturbs the pre-recombination expansion history or initial conditions~\citep[e.g.][]{Abdalla2022} naturally predicts knock-on effects in non-linear structure formation at high redshifts. Such scenarios are especially intriguing because they can either provide a coherent explanation for the observational findings summarized above or, conversely, open a new avenue for constraining early-universe physics with emerging observations of high-redshift galaxies. Some examples are: additional relativistic degrees of freedom~\citep[e.g.][]{Bashinsky2004,Hou2013,Aloni2022}, dark matter-dark radiation interactions that imprint dark-acoustic oscillations~\citep[e.g.][]{Cyr-Racine2013,Shen2024c}, features of the primordial density fields seeded through inflation~\citep[e.g.][]{Chen2010,Achucarro2022}, and early dark energy~\citep[EDE; e.g.][]{Karwal2016, Poulin2018, Poulin2019, Smith2020}. 

In this work, we choose EDE as a representative case study of this broader class. EDE is a transient component active before recombination with an equation-of-state similar to dark energy (before a certain point). This model was originally introduced to address the Hubble tension --- a long-standing discrepancy between the Hubble constant inferred from the Cosmic Microwave Background~\citep[CMB; e.g.][]{Planck2020} and that measured from the local universe~\citep[e.g.][]{Riess2022}. The presence of EDE reduces the physical sound horizon imprinted in the CMB, which in turn requires a shorter distance to the last scattering surface and a higher value of $H_0$ to match observations~\citep[e.g.][]{Poulin2023}. To be specific, we will study the EDE model with parameters constrained in \citet{Smith2022}, based on ACT DR4-era data and chosen for its remarkable success in resolving the Hubble tension. More recent constraints from ACT, Planck, and DESI will be discussed later in Section~\ref{sec:ede}.

Because EDE decays rapidly after recombination and becomes negligible at late times, its impact on cosmic structure formation is indirect, manifesting through shifts in cosmological parameters. Specifically, EDE models tend to increase the primordial fluctuation amplitude ($A_{\rm s}$), spectral index ($n_{\rm s}$), and physical matter density ($\omega_{\rm cdm}$), all of which promote earlier structure growth~\citep{Klypin2021, BK2023, Shen2024b}. In our earlier study~\citep{Shen2024b}, we investigated how this accelerated growth affects UV luminosity and stellar mass functions using a simple empirical model and found that EDE can ease the tension with JWST observations, subject to uncertainties in galaxy formation physics. In this work, we follow up on that idea using large-volume cosmological hydrodynamic simulations employing the IllustrisTNG galaxy formation model, which has been successful in reproducing a wide range of low-redshift galaxy properties~\citep[e.g.][]{Springel2018, Nelson2018, Genel2018, Pillepich2018b, Naiman2018}. This allows us to make more robust predictions for high-redshift galaxy populations while preserving the empirical success of IllustrisTNG at low redshift. A key objective is to assess whether galaxy properties in EDE naturally converge to $\Lambda$CDM predictions at late times. We will also go beyond \citet{Shen2024b} by investigating the formation of disky and quenched galaxies, two populations that are likely to emerge earlier in an EDE cosmology alongside the rapid assembly of massive galaxies and supermassive black holes (SMBHs).

This paper is organized as follows: In Section~\ref{sec:methods}, we introduce the EDE model, the setup of the cosmological hydrodynamic simulations, and an empirical galaxy formation model used for comparison. In Section~\ref{sec:uvlf} and Section~\ref{sec:smf}, we show the predictions of galaxy UV luminosity functions and stellar mass functions in the standard $\Lambda$CDM and the EDE models. In Section~\ref{sec:disk} and Section~\ref{sec:quenched}, we study the evolution of disky galaxies and quenched galaxies in the two models. Discussions and conclusions are presented in Section~\ref{sec:conclusion}.

\begin{figure}
    \centering
    \includegraphics[width=\linewidth]{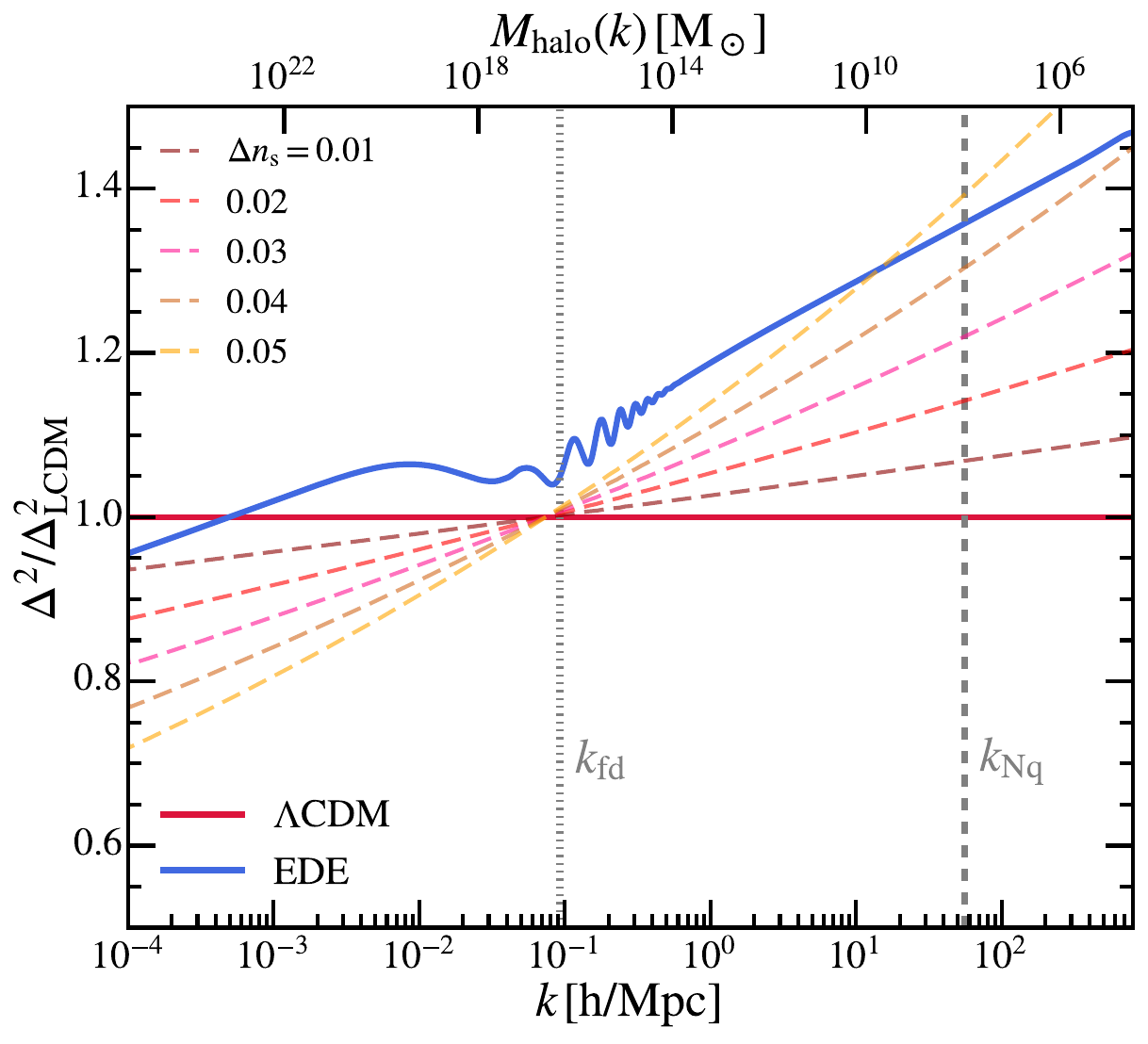}
    \caption{Ratios between the dimensionless linear matter power spectrum ($\Delta^2$) of the EDE and $\Lambda$CDM models. The vertical dotted line indicates the fundamental mode of the box size ($k_{\rm fd}\equiv 2\pi/L_{\rm box}$) in $\Lambda$CDM. The vertical dashed line shows roughly the Nyquist limit of our simulation, $k_{\rm Nq}\equiv (2\pi/L_{\rm box})\, (N^{1/3}_{\rm dm}/2)$, where $N^{1/3}_{\rm dm}$ is the number of dark matter particles per dimension. The top axis shows the halo mass at $z=0$ that corresponds to the $k$-mode~\citep[e.g.][]{Bullock2017}. To guide the eye, we show in dashed lines results with $\Delta n_{\rm s}=0.01,0.02,0.03,0.04,0.05$ compared to $\Lambda$CDM, keeping all other parameters fixed (including $A_{\rm s}$, the normalization at $0.05\Mpc^{-1}$). To the leading order, the shape of the high-k mode power spectrum is mainly set by $n_{\rm s}$, while the normalization is further affected by changes in $A_{\rm s}$, $H_0$, and the transfer function. Meanwhile, the decay of gravitational potential in the EDE case boosts the amplitude of baryon acoustic oscillations, with their peaks shifting to slightly smaller wavenumbers as well.}
    \label{fig:powerspectrum}
\end{figure}

\begin{figure*}
    \raggedright
    \includegraphics[width=0.499\linewidth]{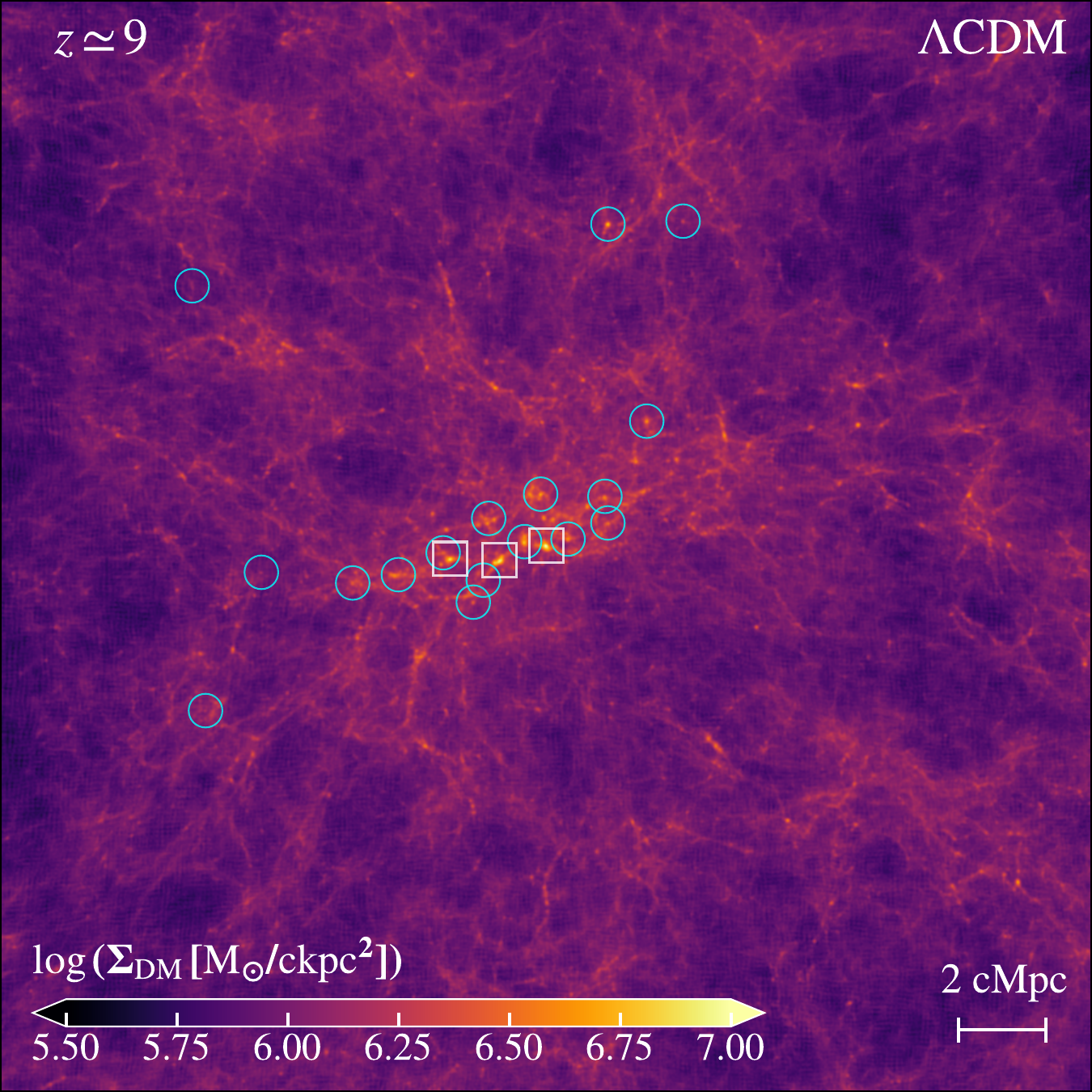}
    \hspace{-0.0055\linewidth}
    \includegraphics[width=0.499\linewidth]{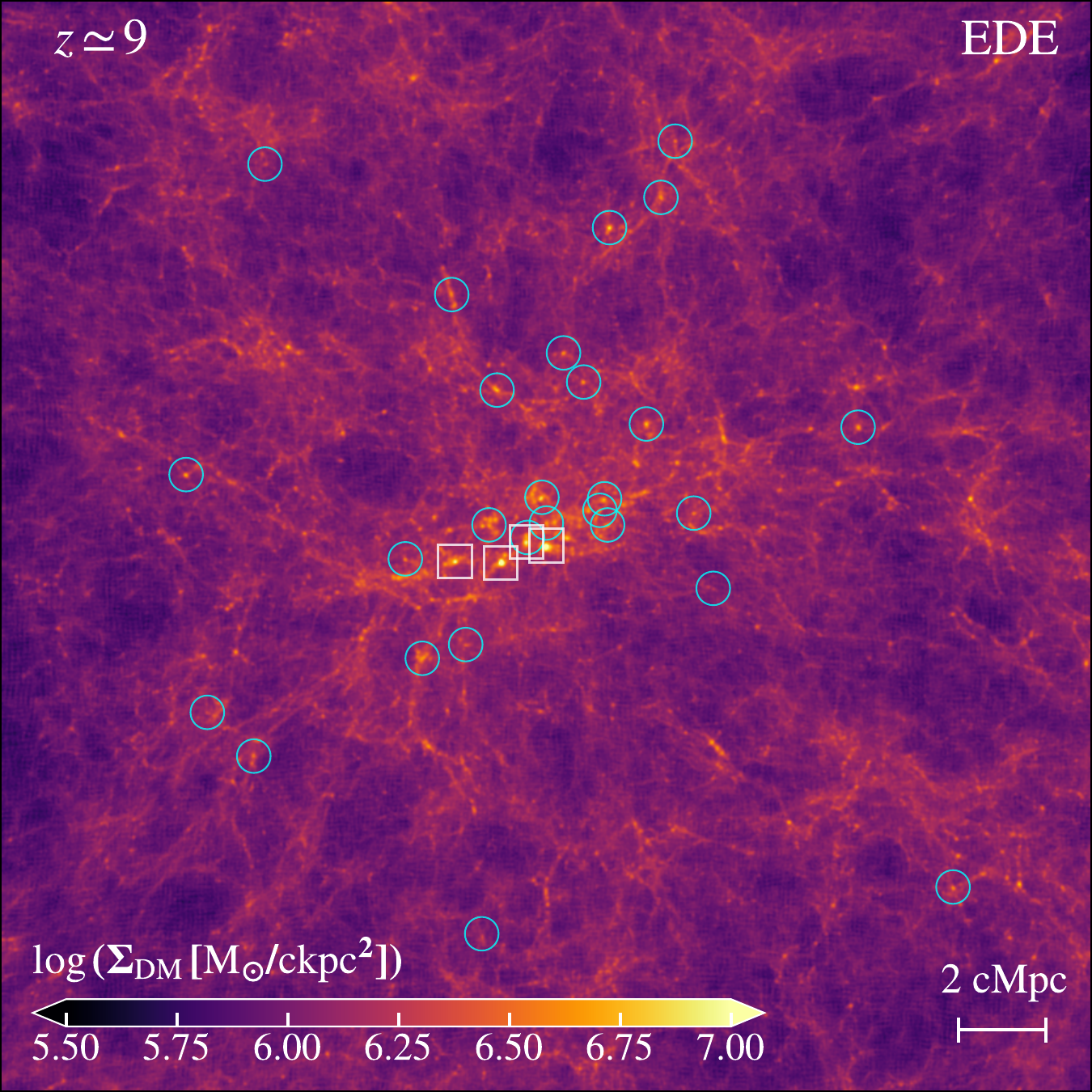}
    \caption{Dark matter surface density in a $25\,{\rm cMpc}$ field of view ($1/4$ of the simulation box length) centered on the most massive halo at $z\simeq 9$. The thickness of the layer for projection is also $25\,{\rm cMpc}$. The left and right panels show the dark matter maps in the $\Lambda$CDM and EDE runs. Dark matter clustering appears stronger in the EDE model. We overlay the location of moderately bright (intrinsic $M_{\rm UV}\leq -18$ mag) and very bright galaxies (intrinsic $M_{\rm UV}\leq -20$ mag) in cyan circles and white squares, respectively. The total number of bright galaxies in this field of view is larger in EDE, reaching $N=31$ compared to $N=20$ in $\Lambda$CDM. However, the clustering power of bright galaxies in EDE is less compared to $\Lambda$CDM due to the lower bias of massive haloes and bright galaxies.}
    \label{fig:img}
\end{figure*}

\begin{figure}
    \centering
    \includegraphics[width=\linewidth]{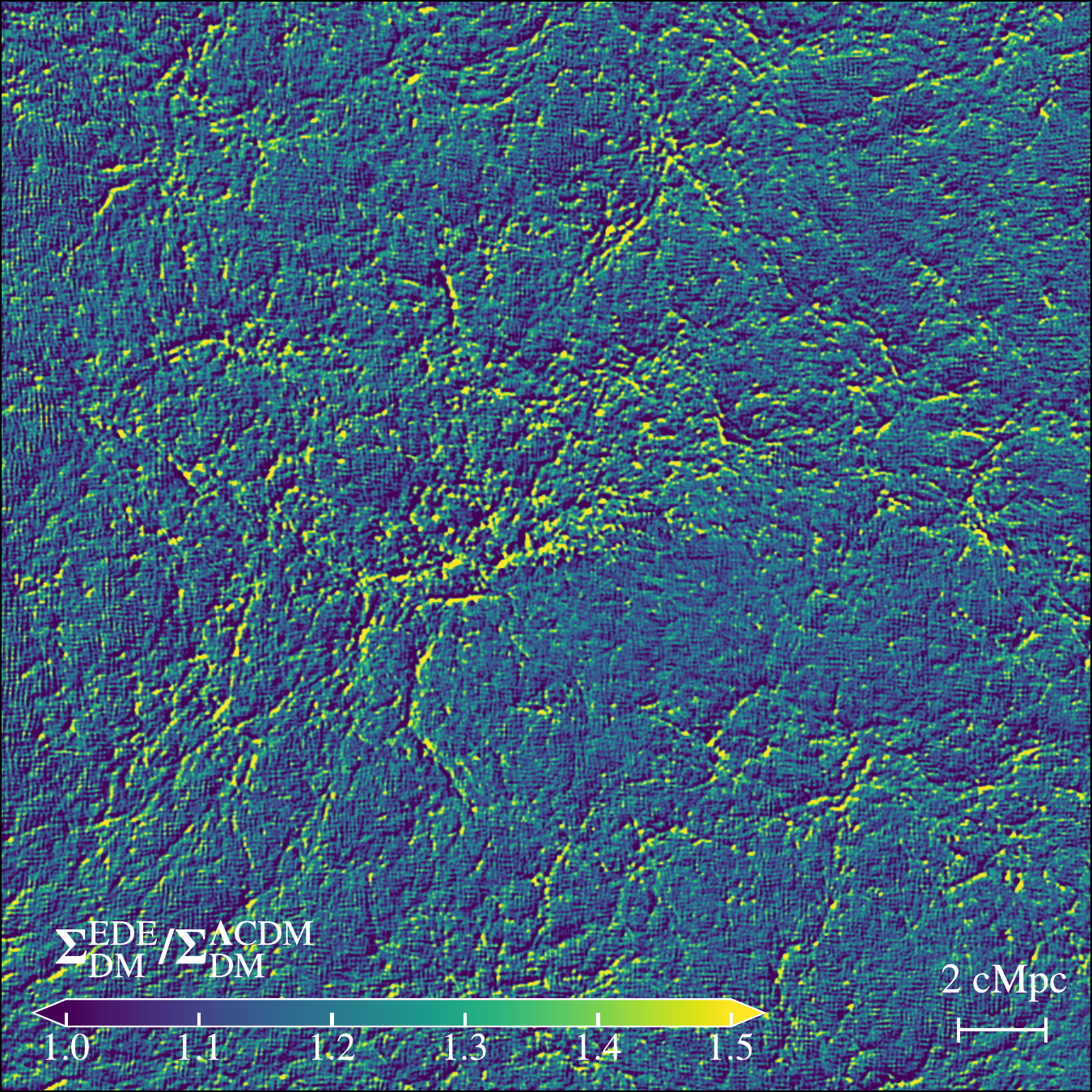}
    \caption{Similar to Figure~\ref{fig:img}, we show the difference of the dark matter surface densities between the EDE and $\Lambda$CDM run. EDE results in enhanced matter clustering in the universe.}
    \label{fig:img2}
\end{figure}

\begin{table}
\centering
\addtolength{\tabcolsep}{8pt}
\def\arraystretch{1.2}
\caption{The best-fit $\pm 1\sigma$ confidence intervals of the cosmological parameters reconstructed in the $\Lambda$CDM and EDE models from the analysis of the ACT DR4 + SPT-3G+Planck TT650TEEE dataset combination~\citep{Smith2022}. $f_{\rm EDE}(z) \equiv \Omega_{\rm ede}(z)/\Omega_{\rm tot}(z)$ is the fraction of energy density contributed by EDE at $z$. $z_{\rm c}$ is the critical redshift when EDE becomes dynamical. $\theta_{\rm i}$ is the initial field value before oscillation of the EDE field. $m$ is the mass and $f$ is the decaying constant of the EDE field, in unit of the Planck scale (Mpl). $H_{\rm 0}$ is the Hubble constant at $z=0$. $\omega_{\rm x} \equiv \Omega_{\rm x}\,h^{2}$, where $h\equiv H_{0}/100$. $A_{\rm s}$ and $n_{\rm s}$ are the normalization and power-law index of the primordial power spectrum. $\Omega_{\rm m}$ is matter density. $S_{8}\equiv \sigma_{8} (\Omega_{\rm m}/0.3)^{1/2}$.}
\begin{tabular}{lll}
Model & $\Lambda$CDM & EDE \\
\hline
\hline
$f_{\rm EDE}(z_{\rm c})$ & - & $0.179^{+0.047}_{-0.04}$ \\
$\log_{10}(z_{\rm c})$ & - & $3.528^{+0.028}_{-0.024}$ \\
$\theta_{\rm i}$ & - & $2.806^{+0.098}_{-0.093}$ \\
$m$ ($10^{-28}$eV) & - & $4.38 \pm 0.49$ \\
$f$ (Mpl) & - & $0.213 \pm 0.035$ \\
\hline
$H_0$ [$\kms\Mpc^{-1}$] & $67.81^{+0.64}_{-0.6}$ & $74.83^{+1.9}_{-2.1}$ \\
$100\omega_{\rm b}$ & $2.249^{+0.014}_{-0.013}$ & $2.278^{+0.018}_{-0.02}$ \\
$\omega_{\rm cdm}$ & $0.1191^{+0.0014}_{-0.0015}$ & $0.1372^{+0.0053}_{-0.0059}$ \\
$10^9A_{\rm s}$ & $2.092^{+0.035}_{-0.033}$ & $2.146^{+0.041}_{-0.04}$ \\
$n_{\rm s}$ & $0.9747^{+0.0046}_{-0.0047}$ & $1.003^{+0.0091}_{-0.0096}$ \\
$S_8$ & $0.821 \pm 0.017$ & $0.829^{+0.017}_{-0.019}$ \\
$\Omega_{\rm m}$ & $0.309^{+0.009}_{-0.008}$ & $0.287 \pm 0.009$ \\
\hline
\end{tabular}
\label{tab:parameters}
\end{table}

\section{Methods}
\label{sec:methods}

\subsection{The early dark energy model}
\label{sec:ede}

Following our previous study~\citep{Shen2024b}, we consider the EDE as a scalar field with an axion-like potential $V_{n}(\phi)\sim [1-{\rm cos}(\phi/f)]^{n}$, where $f$ is the decay constant of the field and $n$ is a constant power-law index. These ultralight axion-like fields can arise generically in string theory~\citep[e.g.][]{Svrcek2006,Arvanitaki2010,Kamionkowski2014,Marsh2016} and have intriguing cosmological implications for dark matter~\citep{Marsh2016} and EDE~\citep{Poulin2019}. The field is frozen at early times and acts as a cosmological constant. The field becomes dynamical at a critical redshift $z_{\rm c}$ as the Hubble friction decreases, eventually settling down around the minimum of the potential and starting oscillation. The effective equation-of-state of the field afterward is $w_{n} \simeq (n-1)/(n+1)$. Here, following \citet{Smith2022}, we consider the $n=3$ case, where the energy density of EDE dilutes faster than radiation. The existence of this additional pre-recombination energy density increases the Hubble parameter around the time of photon decoupling, and reduces the physical sound horizon if $H_0$ is fixed. Therefore, the inferred $H_0$ from the same CMB sound horizon measurements will increase with EDE. Since the energy density of the EDE field dilutes rapidly, it has no direct impact on structure formation, but its effects are indirectly imprinted through changes in cosmological parameters and transfer function. We choose EDE and cosmological parameters as the best-fit values in \citet{Smith2022}, which are constrained by joint data from the ACT, SPT, and Planck. They are summarized in Table~\ref{tab:parameters} along with the best-fit parameters in the standard $\Lambda$CDM cosmology. The best-fit $H_0$ in the EDE cosmology is around $73-74\kms \Mpc^{-1}$, in agreement with the local constraints~\citep[e.g.][but see also e.g. \citealt{Freedman2019,Freedman2025}]{Riess2022,Riess2024,Brout2022}. The transfer functions and linear matter power spectra are calculated using the Code for Anisotropies in the Microwave Background (\textsc{Camb}; \citealt{CAMB1,CAMB2}) and specifically the early quintessence model~\citep{Smith2020} implemented there. We have verified that alternative codes such as \textsc{Axiclass}~\citep[e.g.][]{Poulin2018} produce identical results. 

In Figure~\ref{fig:powerspectrum}, we compare the linear matter power spectra of the EDE and standard $\Lambda$CDM cosmologies. The EDE cosmology exhibits enhanced small-scale power, primarily driven by the larger primordial scalar spectral index $n_{\rm s}$. To illustrate this, we include reference lines corresponding to $\Delta n_{\rm s} = 0.01-0.05$ in the figure. The line with $\Delta n_{\rm s} = 0.03 \approx n^{\rm EDE}_{\rm s} - n^{\Lambda{\rm CDM}}_{\rm s}$ matches well the slope of the EDE power spectrum. Additional differences in the normalization arise from a combination of a higher $A_{\rm s}$, a suppressed transfer function, and an increased $H_0$ in EDE. This boost in the small-scale power spectrum is a key factor in accelerating early structure formation. To isolate this effect, in Figure~\ref{fig:hmf_z12} we compare the halo mass functions in $\Lambda$CDM, EDE, and two intermediate models: one with only $H_0$ increased and another with only the power spectrum modified (keeping all other parameters the same as the $\Lambda$CDM values). The ``power spectrum-only'' model reproduces the EDE halo mass function at $z \simeq 12$, confirming that the change in the small-scale power spectrum is the dominant driver of accelerated halo formation in EDE. 

The fitting results from \citet{Smith2022}, which we adopt here, were derived using ACT DR4, SPT-3G 2018, and Planck data. These datasets hinted at a preference for EDE over $\Lambda$CDM (see also e.g. \citealt{Hill2022}) while simultaneously solving the Hubble tension. However, subsequent analyses incorporating Planck NPIPE PR4 and BOSS BAO data significantly weakened the support for this scenario~\citep{Efstathiou2024}. More recently, ACT DR6 data alone placed moderately relaxed constraints on EDE, with $f_{\rm EDE} \lesssim 0.12$ at the 95\% confidence level. Despite this, an updated joint analysis using ACT DR6, Planck NPIPE, and DESI DR2 found that an EDE model with $f_{\rm EDE} \sim 0.13$ and $H_0 \sim 72.5~\kms,{\rm Mpc}^{-1}$ remains statistically favored over $\Lambda$CDM, reducing the Hubble tension to $<2\sigma$ level~\citep{Poulin2025}. These updated EDE models are less aggressive than the one studied here (which aims to fully resolve the Hubble tension), but the associated cosmological parameters relevant to the power spectrum (e.g. $n_{\rm s} \sim 1$) are similar. We emphasize that the enhanced small-scale power and accelerated early structure formation are generic predictions of beyond-$\Lambda$CDM models that increase the pre-recombination expansion rate. Such scenarios increase the angular scale of Silk damping and require a more tilted primordial spectrum to match CMB observations. In this paper, we aim to complete the connection between this class of early-universe physics and high-redshift galaxy formation, and the implications extend beyond the EDE scenario alone.

\begin{table*}
    \addtolength{\tabcolsep}{4pt}
    \renewcommand{\arraystretch}{1.05}
    \centering
    \caption{ \textbf{A summary of the simulations. Each column corresponds to the following information:} \newline \hspace{\textwidth}
    (\textbf{1}) Name of the simulation. 
    (\textbf{2}) $L_{\rm box}$: Side-length of the periodic simulation box.  
    (\textbf{3}) $N_{\rm part}$: Number of particles (cells) in the simulation. In the initial conditions, there are an equal number of dark matter particles and gas cells. 
    (\textbf{4}) $z_{\rm f}$: Lowest redshift reached by the simulation.
    (\textbf{5}) $m_{\rm dm}$: Mass of dark matter particles, which is constant over time. 
    (\textbf{6}) $m_{\rm b}$: Mass of gas cells in the initial conditions as a reference for the baryonic mass resolution. The gas cells are (de-)refined so that the gas mass in each cell is within a factor of two of this target gas mass. Stellar particles stochastically generated out of gas cells have initial masses typically comparable to $m_{\rm b}$ and are subject to mass loss via stellar evolution~\citep{Vogelsberger2013}. 
    (\textbf{7}) $\epsilon$: The comoving (Plummer-equivalent) gravitational softening length for the dark matter and stellar particles. The minimum gravitational softening length of gas cells (adaptively softened) is $1/4$ of this. 
    (\textbf{8}) $\epsilon^{z=6}_{\rm phy}$: The physical gravitational softening length at $z=6$ for reference. 
    (\textbf{9}) $r^{\rm min}_{\rm cell}$: The minimum physical size of gas cells at the end of the simulations.
    }
    \begin{tabular}{lcccccccccc}%
        Simulation Name & $L_{\rm box}\,[{\rm cMpc}]$ & $N_{\rm part}$ & $z_{\rm f}$ & $m_{\rm dm}\,[\msun]$ & $m_{\rm b}\,[\msun]$ & $\epsilon\,[{\rm ckpc}]$ & $\epsilon^{z=6}_{\rm phy}\,[{\rm pc}]$ & $r^{\rm min}_{\rm cell}\,[{\rm pc}]$ \\
        \hline
        \hline
        $\Lambda$CDM & 100 & $2\times 1200^{3}$ & 3 & $1.9\times 10^7$ & $3.6\times 10^6$ & 1.2 & 171 & 17.8 \\
        EDE & 100 & $2\times 1200^{3}$ & 3 & $2.2\times 10^7$ & $3.7\times 10^6$ & 1.2 & 171 & 18.3 \\
        \hline
    \end{tabular}
    \label{tab:sims}
    \renewcommand{\arraystretch}{0.9090909090909090909}
\end{table*}

\subsection{Simulations}
\label{sec:sim}

We perform a suite of cosmological hydrodynamic simulations utilizing the moving-mesh multi-physics code \textsc{Arepo}~\citep{Springel2010, Pakmor2016, Weinberger2020}. Gravity is solved using the hybrid Tree-Particle Mesh (PM) method \citep{Barnes1986}. The hydrodynamics is solved using the quasi-Lagrangian Godunov method \citep{Godunov1959} on an unstructured Voronoi mesh grid (see e.g. \citealt{Springel2010} and \citealt{Vogelsberger2020NatR} for a review). The simulations employ the IllustrisTNG galaxy formation model~\citep{Weinberger2017,Pillepich2018}, which is an update of the Illustris model \citep{Vogelsberger2014,Vogelsberger2014b}. The simulations include: (1) primordial and metal-line cooling of gas~\citep{Smith2008,Wiersma2009} in the presence of a redshift-dependent, spatially uniform UV radiation background~\citep{FG2009} with corrections for self-shielding~\citep{Katz1992,Rahmati2013}, (2) a two-phase effective equation-of-state model for the interstellar medium (ISM) and pressurization from stellar feedback at the sub-resolution level~\citep{Springel2003}, (3) star-formation in dense ($n_{\rm H}\geq 0.1\,{\rm cm}^{-3}$) gas following the empirically defined Kennicutt–Schmidt relation and stellar particles are spawn in a stochastic fashion representing single stellar populations with the \citet{Chabrier2003} stellar IMF, (4) metal enrichment from stellar evolution in the asymptotic giant branch (AGB) phase and supernovae (SNe), (5) galactic winds driven by SNe implemented with a non-local thermal plus kinetic wind scheme, and wind particle energies and velocities are prescribed to depend on the SNe rates (from stellar evolution model), the local dark matter velocity dispersion, redshift, and gas metallicity, and (6) SMBH formation, growth, and feedback in the thermal and kinetic modes as described in \citet{Weinberger2017}. The model has been extensively tested in large-scale cosmological simulations and can produce realistic galaxy populations that match a wide range of observations, in terms of large-scale galaxy clustering~\citep[e.g.][]{Springel2018}, galaxy properties~\citep[e.g.][]{Nelson2018, Genel2018, Pillepich2018b, Naiman2018}, especially at moderately high redshifts~\citep[e.g.][]{Vogelsberger2020, Shen2020, Shen2022, Kannan2023}. In the following analysis, we will evaluate when the EDE model predictions converge to the standard $\Lambda$CDM and validate whether the success of the IllustrisTNG galaxy formation model at lower redshifts is preserved.

The initial conditions of the simulations are generated with the \textsc{Gadget4} code~\citep{Springel2021} using the second-order Lagrangian perturbation theory at the initial redshift of $z_{\rm i}=99$. The cosmological parameters used are shown in Table~\ref{tab:parameters}, and the linear matter power spectra are self-consistently derived as introduced in the previous section. In the initial conditions, the gas perfectly follows the dark matter distribution and is assumed to have primordial composition with hydrogen and helium mass fractions of $X=0.76$ and $Y=0.24$, respectively. The simulations and the associated numeric parameters are summarized in Table~\ref{tab:sims}. The halo mass of a galaxy is defined as the sum of the mass of all particles gravitationally bound to the subhalo identified by \textsc{Subfind-hbt} \citep{Springel2001,Springel2021}. The stellar mass of a galaxy is defined as the sum of the current mass of all stellar particles within twice the stellar half-mass radius ($r^{\ast}_{\rm 1/2}$). The star-formation rate (SFR) of a galaxy is defined as the sum of instantaneous SFRs of gas cells in the same aperture. 

\subsection{Observables for model-data comparison}

\noindent \textbf{UV luminosities of galaxies}:
We use the code \textsc{Synthesizer}\footnote{\href{https://github.com/synthesizer-project/synthesizer}{https://github.com/synthesizer-project/synthesizer}}~\citep{Lovell2024-syn,Roper2025-syn} to model the intrinsic emission from simulated galaxies. Specifically, we adopt the stellar spectra (including binary stars) from \textsc{BPASS}-v2.2.1 \citep{Eldridge2017,Stanway2018} to model the rest-frame UV and optical spectral energy distributions (SEDs) of galaxies. We assume the \citet{Chabrier2003} IMF covering $0.1$ to $300\msun$. The spectra have been post-processed with the photoionization code \textsc{Cloudy}-v23.01~\citep{Ferland1998,Chatzikos2023} to obtain nebular continuum and line emission. The hydrogen density of the nebula is assumed to be $n_{\rm H}=1000\,{\rm cm}^{-3}$ and the ionization parameter is set to $U=0.01$~\citep{Wilkins2020}, broadly consistent with the ionized gas properties inferred for high-redshift galaxies from recent JWST observations~\citep[e.g.][]{Reddy2023,Calabro2024,Topping2024b,Naidu2025}. For each star particle, we take the combination of stellar age and initial metallicity to evaluate the spectral emissivity, and they are aggregated (with stellar initial masses as weights) to obtain the galaxy SED. We measure the rest-frame UV flux of a galaxy using a top-hat filter centered at $1500$~\AA\, with a width $100$~\AA. For our predictions of UV luminosities at $z\geq 10$, we assume no dust attenuation, motivated by the observed blue UV slopes of galaxies at those redshifts~\citep[e.g.][]{Cullen2024,Roberts-Borsani2024,Topping2024}. At $z<10$, we use the empirical approach as will be described in Section~\ref{sec:empirical} below. \\

\noindent \textbf{Classification of star-forming and quenched galaxies}: 
To separate star-forming and quiescent galaxies, we use a time-dependent cut in specific star-formation rate (sSFR), ${\rm sSFR} = 0.2/t_{\rm H}(z)$, where $t_{\rm H}(z)$ is the Hubble time at $z$. This cut has been widely applied in the literature~\citep[e.g.][]{Franx2008, Gallazzi2014, Pacifici2016, Tacchella2022,Carnall2023b}, and yields results that are broadly consistent with the classical UVJ selection~\citep[e.g.][]{Schreiber2018,Carnall2018,Leja2019,Carnall2023b}. \\

\noindent \textbf{``Diskiness''}:
We define the face-on direction of a galaxy as the direction of the total angular momentum of stellar particles within twice $r^{\ast}_{\rm 1/2}$. We decompose the stellar content of a galaxy by the circularity parameters of stellar particles, $j_z/j_{\rm tot}$, where $j_z$ is the angular momentum of the particle aligned with the galaxy face-on vector and $j_{\rm tot}$ is the total angular momentum of this particle. Following \citet{Tacchella2019}, we define that a stellar particle belongs to the disk if $j_z/j_{\rm tot}>0.7$. We note that a purely isotropic system will still contain $\simeq 15\%$ of orbits with $j_z/j_{\rm tot}>0.7$ according to this definition. The stellar disk fraction is the mass fraction of the disk component to the total stellar mass. The gas disk fraction is defined in the same way as the stellar content, except that the face-on direction is instead defined by the total angular momentum of gas cells within twice $r^{\ast}_{\rm 1/2}$. A galaxy is considered to have a stellar disk if the D/T (disk-to-total mass ratio) of stars exceeds $0.7$. This is roughly consistent with the typical D/T values of optical blue, star-forming galaxies at around the Milky Way mass at $z=0$ identified in the IllustrisTNG simulations~\citep[e.g.][]{Tacchella2019,Zana2022}. A galaxy is considered to have a gaseous disk if the D/T of gas exceeds $0.8$. In Figure~\ref{fig:fdisk_z6}, we illustrate the distribution of simulated galaxies in the D/T$-M_{\ast}$ plane and the cuts we apply. 

\begin{figure}
    \centering
    \includegraphics[width=\linewidth]{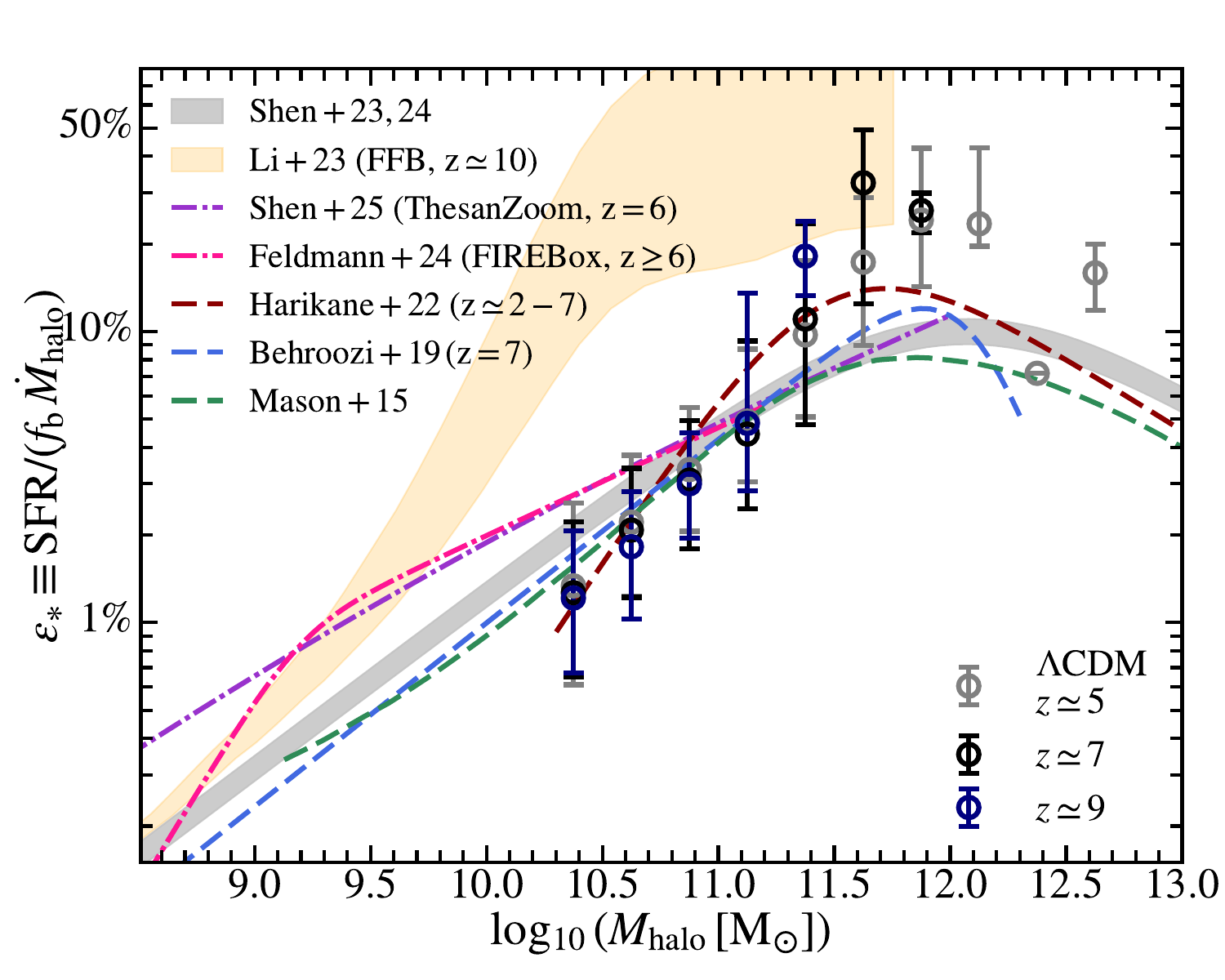}
    \caption{Star-formation efficiency (SFE) versus halo mass. We show our $\Lambda$CDM simulation results as open circles with $1\sigma$ error bars at $z\simeq 5,7$, and $9$. The EDE run yields identical results. We compare them to the SFE assumed in our empirical model~\citep{Shen2023,Shen2024b} and other choices in literature~\citep[][]{Mason2015,Behroozi2019,Harikane2022,Li2023,Feldmann2024,Shen2025} as labelled. The SFEs from our simulations (as the prediction of the IllustrisTNG model) show almost no redshift dependence. They are consistent with canonical models until reaching $M_{\rm halo}\sim 10^{12}\msun$, above which AGN feedback quenches star-formation. By construction, the model does not predict a feedback-free starburst (FFB) regime as in \citet{Dekel2023} and \citet{Li2023}.}
    \label{fig:sfe}
\end{figure}

\begin{figure*}
    \raggedright
    \includegraphics[width=0.49\linewidth, trim={0 0.2cm 0 0.2cm}]{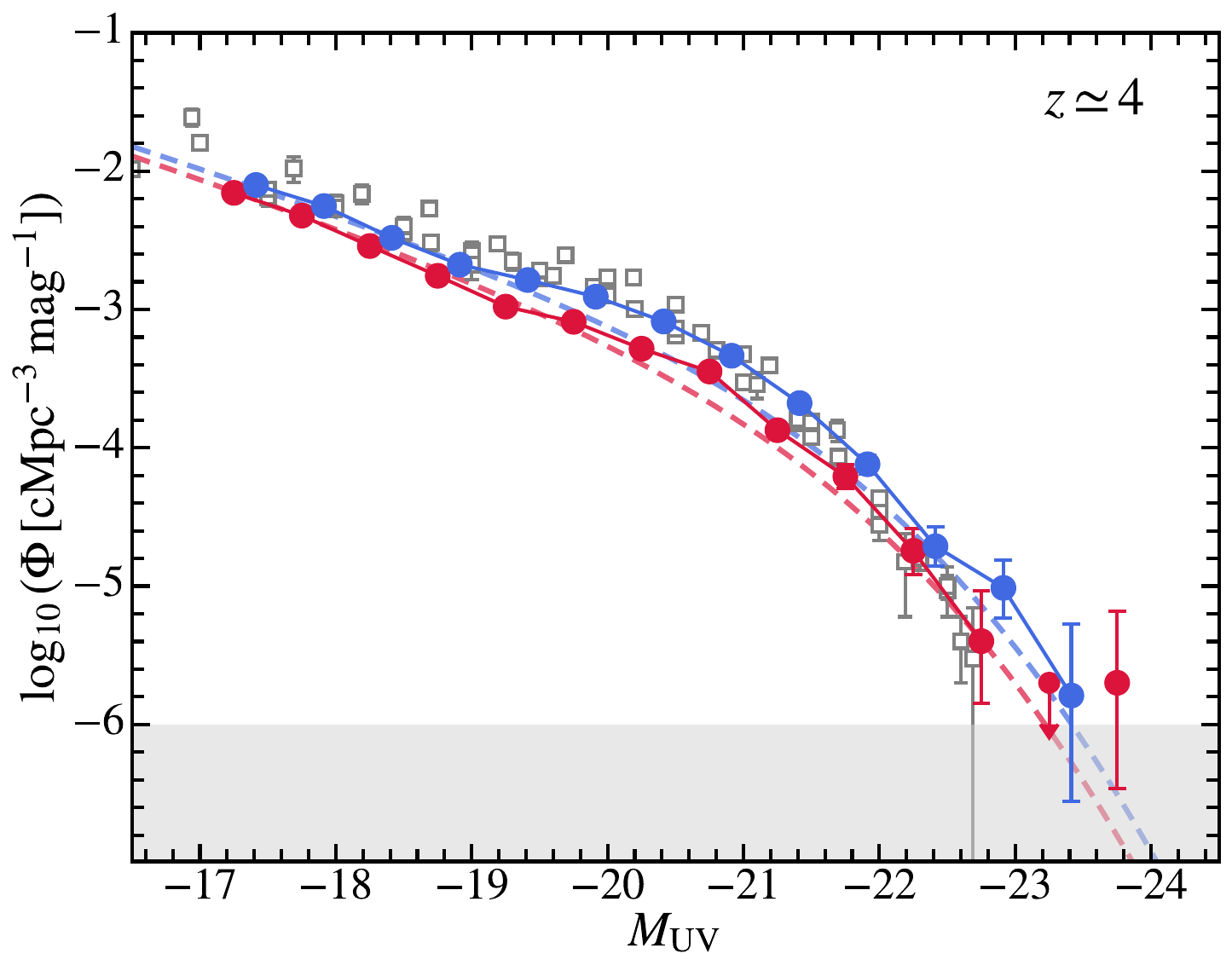}
    \includegraphics[width=0.49\linewidth, trim={0 0.2cm 0 0.2cm}]{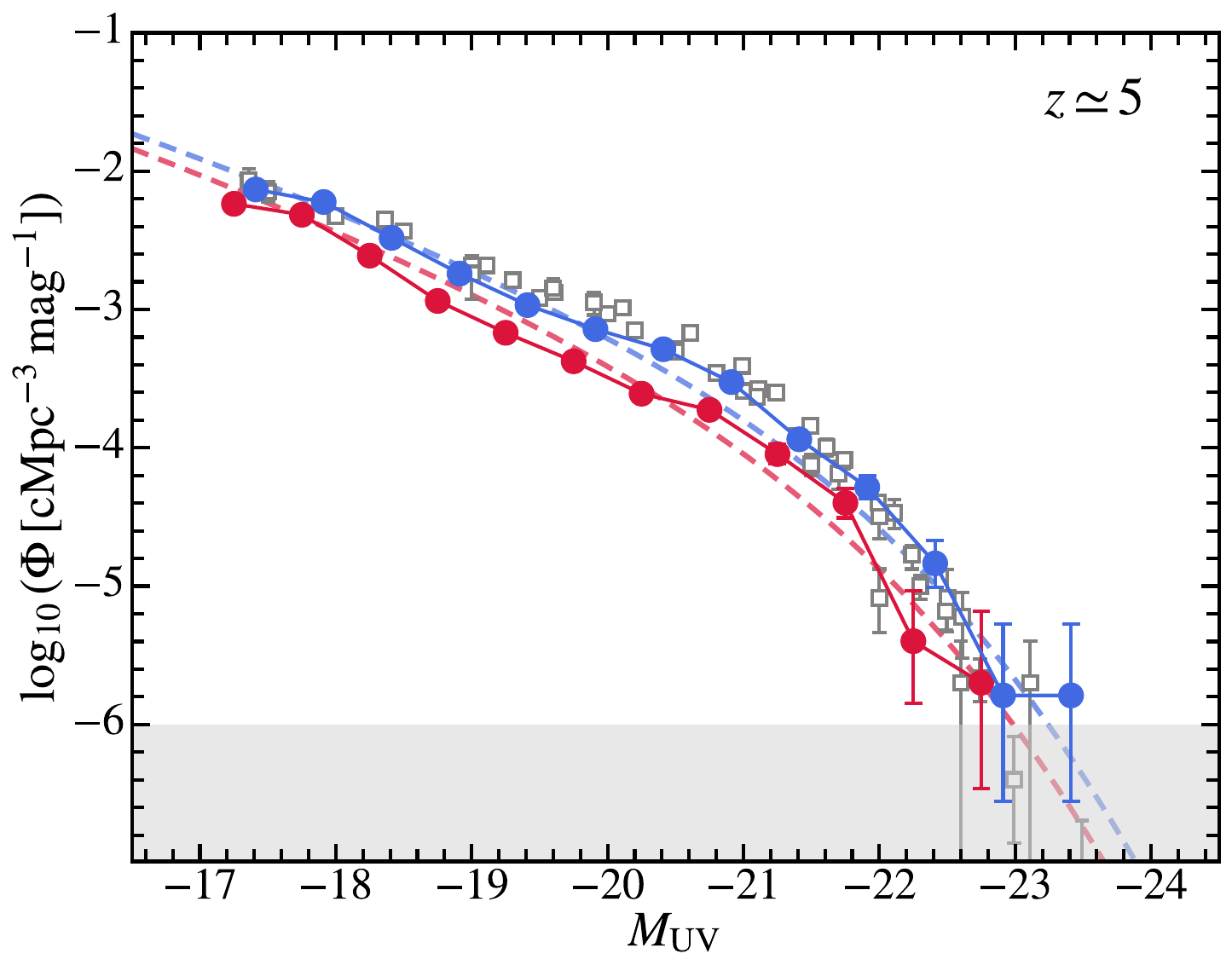}
    \includegraphics[width=0.49\linewidth, trim={0 0.2cm 0 0.2cm}]{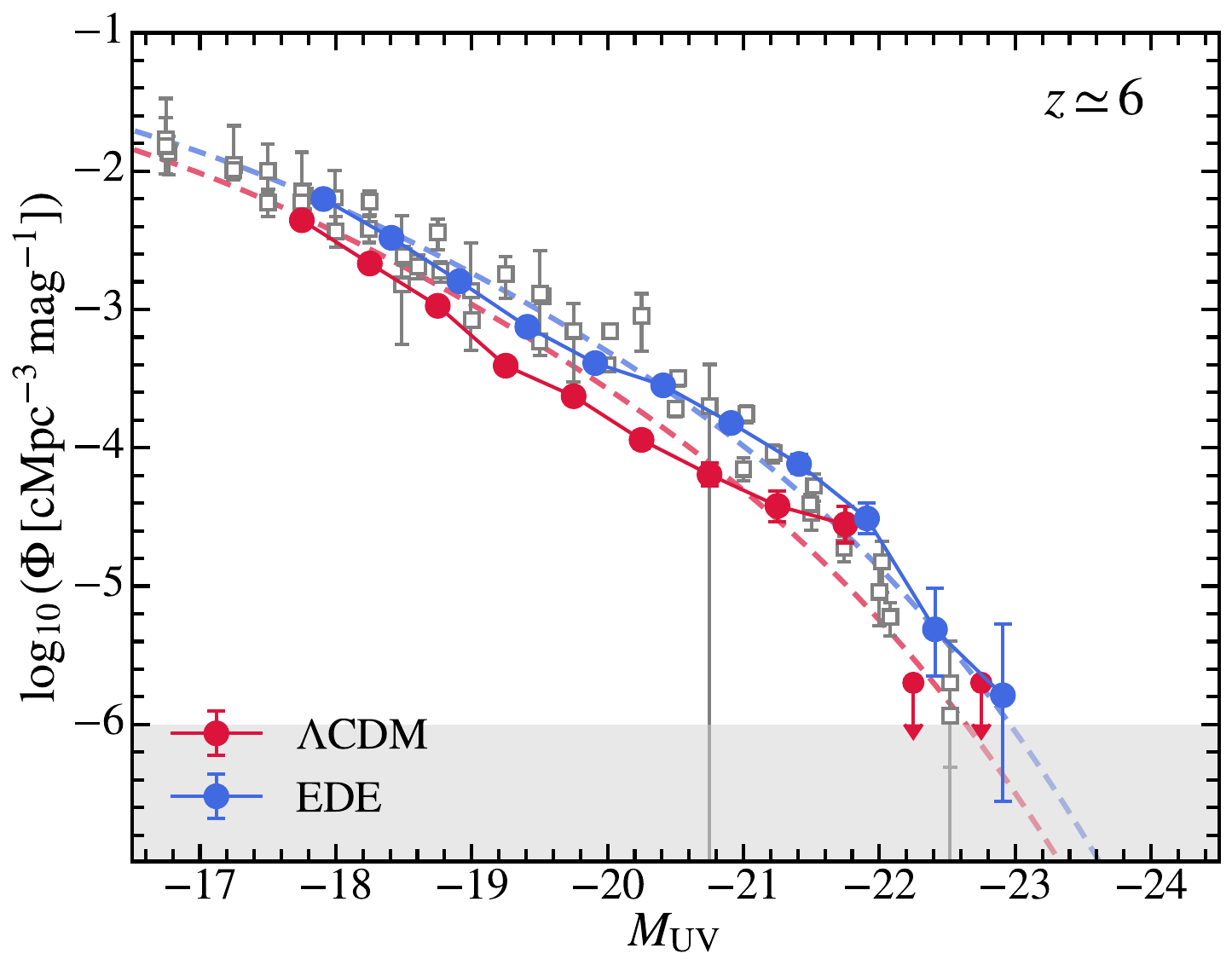}
    \includegraphics[width=0.49\linewidth, trim={0 0.2cm 0 0.2cm}]{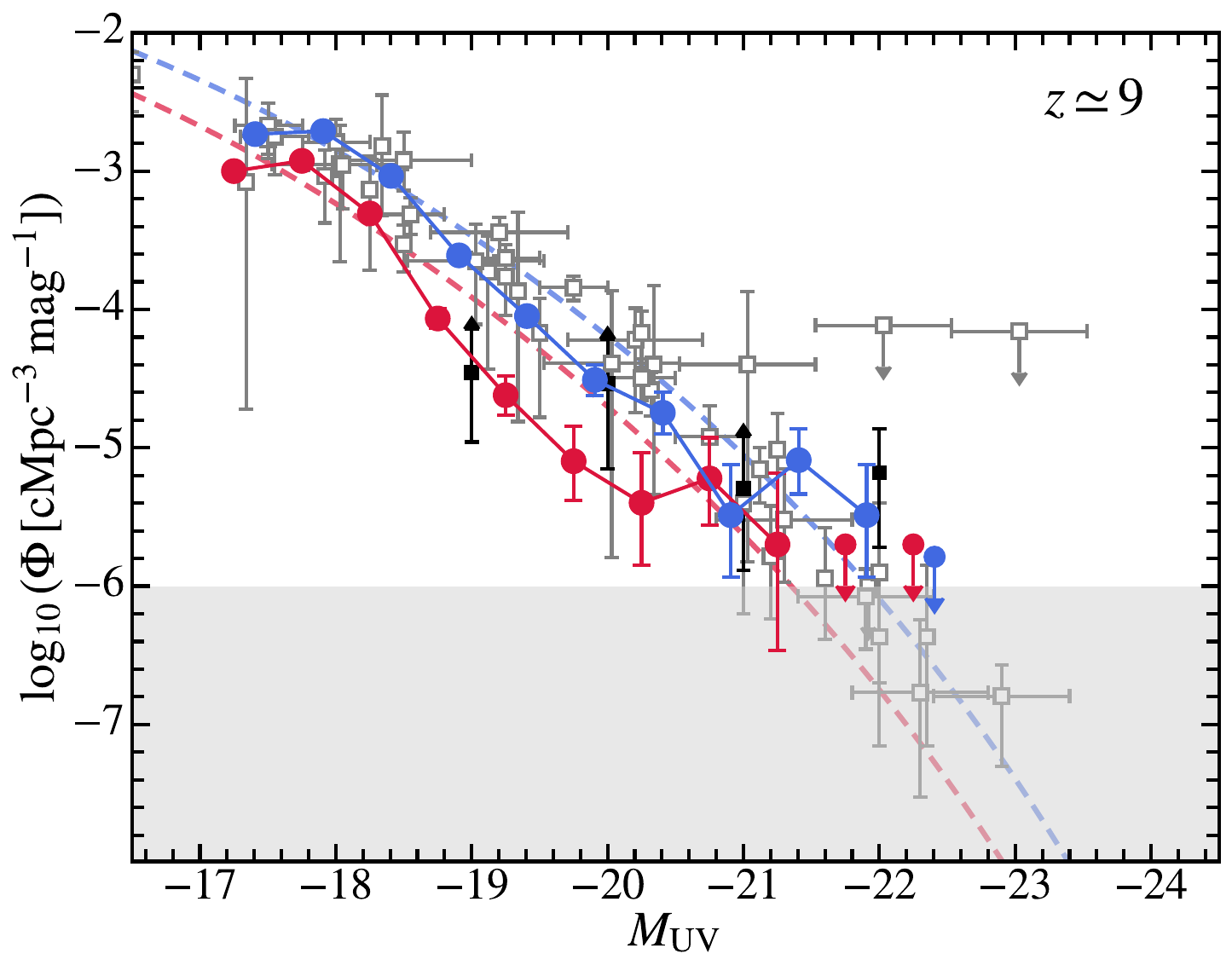}
    \includegraphics[width=0.49\linewidth, trim={0 0.2cm 0 0.2cm}]{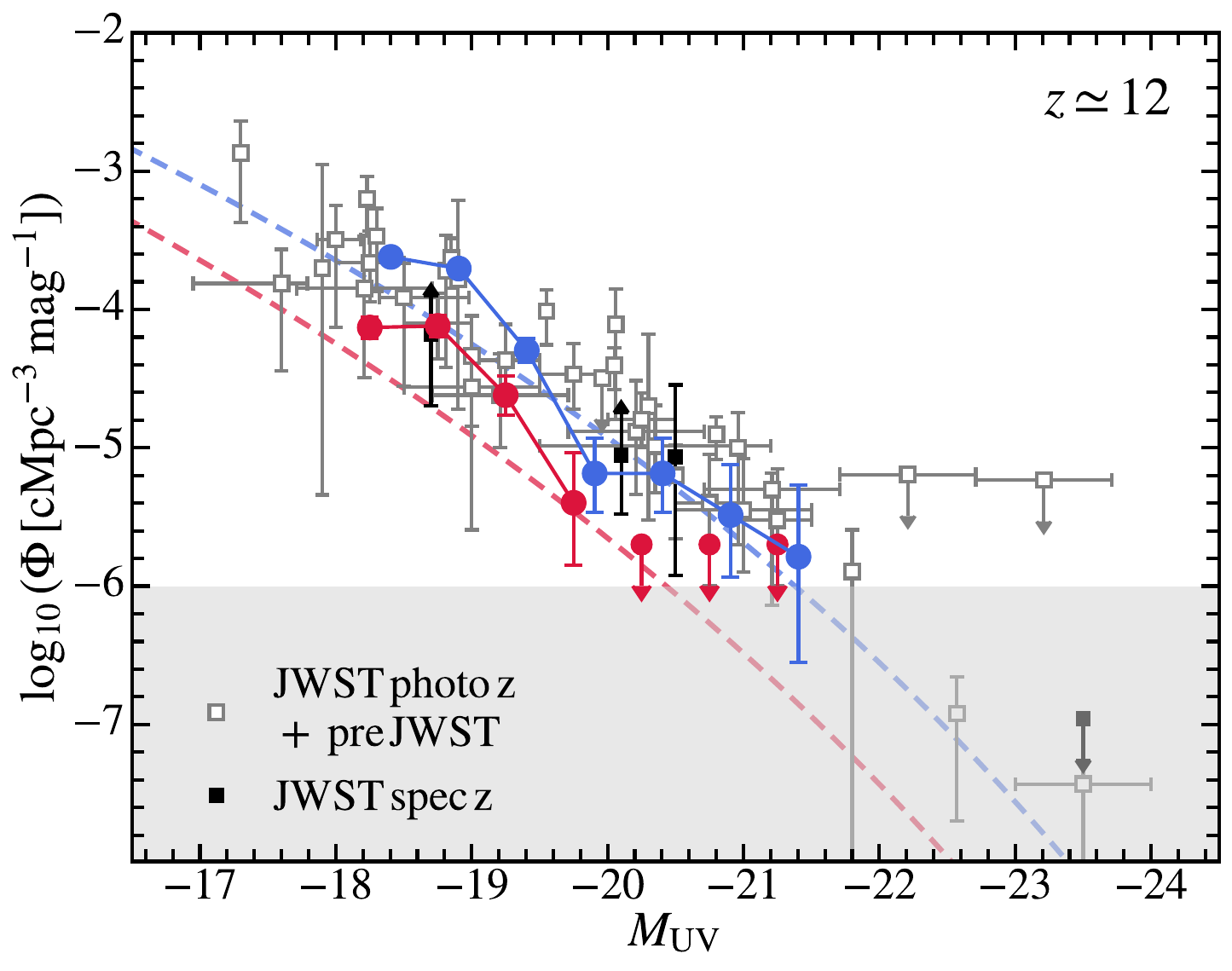}
    \includegraphics[width=0.49\linewidth, trim={0 0.2cm 0 0.2cm}]{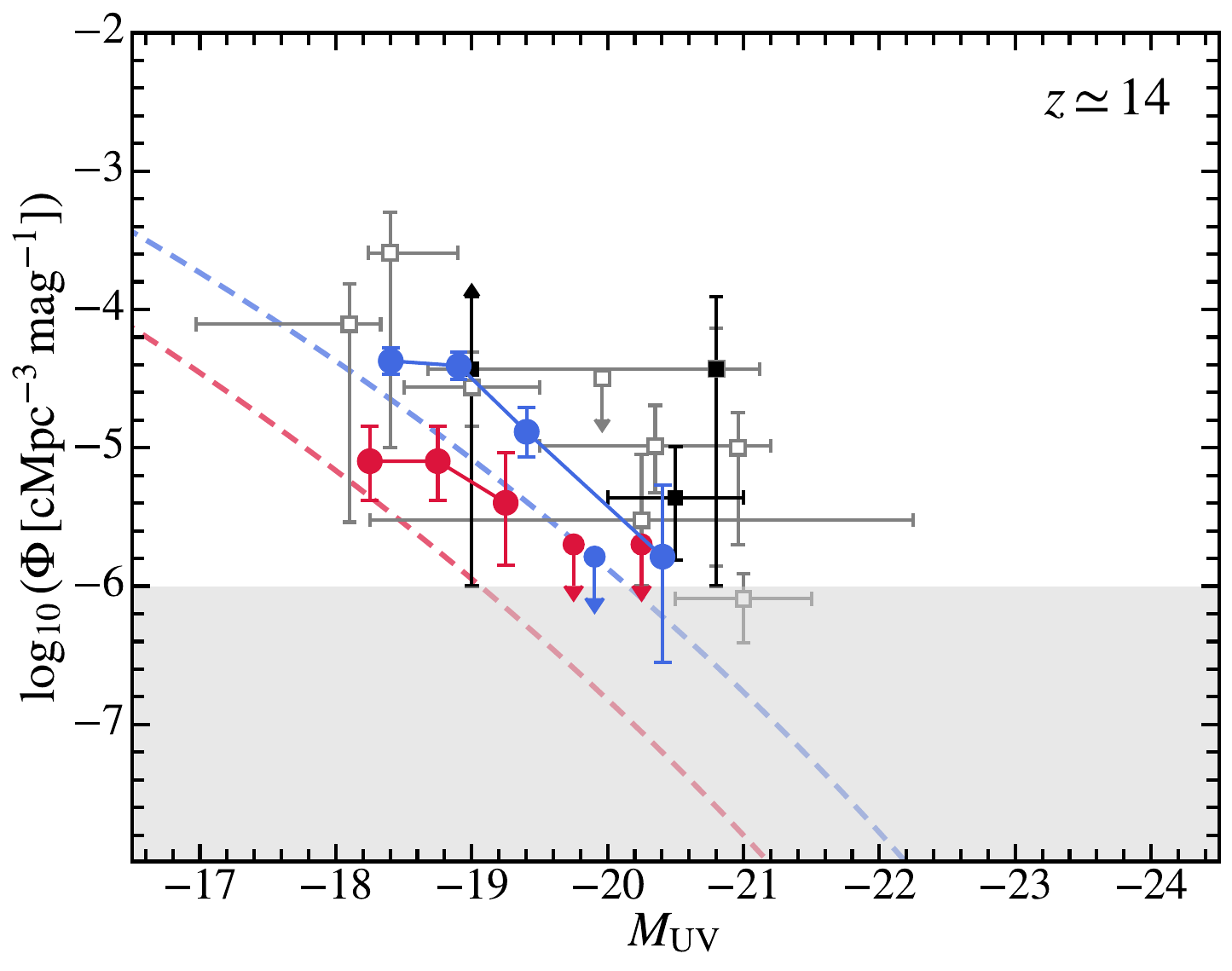}
    \caption{Galaxy rest-frame UV luminosity function at $z\simeq 4-14$ in the $\Lambda$CDM and EDE simulations. Binned estimations from the simulations are shown with solid circles, and Poisson errors are computed using the formula in \citet{Gehrels1986}. Dust attenuation has been applied at $z<10$ as described in Section~\ref{sec:empirical}. We compare the results to the observational constraints compiled in \citet{Shen2024b}. The predictions from the empirical model~\citep{Shen2023,Shen2024b} are shown in dashed lines for comparison. The shaded region in each panel indicates the ``one galaxy per mag'' limit of our simulation volume. The $\Lambda$CDM run underpredicts the UV luminosity function. In particular, at $z>10$ where no dust attenuation has been applied, the UV luminosity functions are still $0.5-1$ dex lower than the observational constraints. On the contrary, over the entire redshift range, the UV luminosity functions from the EDE run agree almost perfectly with observations.}
    \label{fig:uvlf}
\end{figure*}

\subsection{An empirical model for comparison}
\label{sec:empirical}

We adopt the empirical model outlined in \citet{Shen2023,Shen2024b} as a baseline for interpreting our simulation results. The goal is to evaluate the performance of this simple empirical approach in capturing the non-linear evolution of structures in the universe, and use it for several volume corrections that will be discussed in Section~\ref{sec:smf}. Here, we provide a summary of the key ingredients in this model. We start from the halo mass functions constructed following Press-Schechter-like theories \citep[e.g.][]{Press1974, Bond1991, Sheth2001} as implemented in the {\sc HMF} code~\citep{hmf2,hmf3}. The definition of halo mass follows the virial criterion in \citet{Bryan1998}. The halo accretion rates ($\dot{M}_{\rm halo}$) are computed using the fitting formulae in \citet{RP2016}, which is calibrated on the Bolshoi-Planck and MultiDark-Planck cosmological simulations~\citep{Klypin2016}\footnote{The dependence of $\dot{M}_{\rm halo}$ on cosmology is absorbed in the $H(z)$ term as found in numeric studies using different cosmological parameters~\citep[e.g.][]{Fakhouri2010,Klypin2016,Dong2022}.}. 

We connect the SFRs of galaxies with their host halo accretion rates as ${\rm SFR} = \varepsilon_{\ast}\,f_{\rm b}\,\dot{M}_{\rm halo}$, where $f_{\rm b}\equiv \Omega_{\rm b}/\Omega_{\rm m}$ is the universal baryon fraction and $\varepsilon_{\ast}$ is the (halo-scale) star-formation efficiency (SFE). We assume $\varepsilon_{\ast}$ to depend only on halo mass as prescribed in \citet{Shen2024b}, which represents a ``median'' representation of pre-JWST galaxy formation models. In Figure~\ref{fig:sfe}, we compare the SFE model in \citet{Shen2024b} with other choices in observational and theoretical literature, including the observational constraints in \citet{Harikane2022} based on a halo occupation distribution, the \textsc{Universe Machine} predictions at $z\simeq 7$, the empirical choice calibrated in \citet{Mason2015}, results from the FIREbox~\citep{Feldmann2024} and the \thesanzoom~\citep{Shen2025,Kannan2025} simulations, and the feedback-free starburst (FFB) scenario at $z\simeq 10$~\citep{Li2023}. We also overlay the SFE from our cosmological simulations, which is derived as the ratio between SFR and the halo accretion rates computed from halo mass using the same fitting formula in \citet{RP2016}. The EDE and $\Lambda$CDM runs yield indistinguishable SFEs, so for brevity we show the $\Lambda$CDM results only. The SFE resulting from the IllustrisTNG model exhibits little redshift evolution and agrees broadly with other empirical inferences up to $M_{\rm halo} \sim 10^{12} \msun$, above which AGN feedback quenches star-formation. The behavior is a consequence of the effective equation-of-state of the ISM and the Kennicutt–Schmidt law assumed in the star-formation recipe. At the low-mass end, we find non-negligible differences relative to simulations with alternative ISM prescriptions~\citep[e.g.][]{Feldmann2024,Shen2025,Wang2025}, which tend to produce moderately higher SFE. Finally, by construction, our simulations and empirical model do not realize the FFB regime.

The intrinsic UV-specific luminosity $L_{\nu}(\mathrm{UV})$ is from SFR, assuming the conversion factor $\kappa_{\rm UV}$ in \citet{Madau2014} and the \citet{Chabrier2003} IMF. Dust attenuation is modeled using a combination of the scaling relations between attenuation in UV ($A_{\rm UV}$), UV continuum slope ($\beta$), and the observed $M_{\rm UV}$. For the first part, we take the function form $A_{\rm UV}=C_0 + C_1\,\beta$~\citep{Meurer1999} and typical values $C_0 \simeq 4.9$, $C_1 \simeq 2$ from recent ALMA measurements~\citep{Bowler2024}. The second part, the $\beta$-$M_{\rm UV}$ relation, is taken from observational measurements in \citet{Bouwens2014} at $z<8$ and \citet{Cullen2023} at $z\geq 8$. The scatter is $\sigma_{\beta} \simeq 0.35$~\citep{Bouwens2014, Rogers2014, Cullen2023}, which can impact the average attenuation at some fixed intrinsic $M_{\rm UV}$ values~\citep[e.g.][]{Tacchella2013,Tacchella2018,Vogelsberger2020}. The same dust attenuation recipe is also applied to our simulation results. 

A convolution of the resulting $M_{\rm UV}-M_{\rm halo}$ relation and the halo mass function is conducted to obtain the UV luminosity function. One difference from \citet{Shen2024b} is that we assume a relatively low and constant UV variability $\sigma_{\rm UV} = 0.75\,\mmag$ (at fixed halo mass) to mimic the low level of burstiness in star-formation in the IllustrisTNG galaxy formation model. The choice is purely for interpretation and comparison to simulation results~\footnote{Although, as a coincidence, this is more consistent with recent observational constraints on the burstiness of star-formation at $z\gtrsim 3$~\citep[e.g.][]{Simmonds2025}.}. 

In this empirical model, we derive the stellar-to-halo mass relation of galaxies by integrating the SFE as
\begin{align}
    M_{\ast}(M_{\rm halo}) = & (1-R)\, \bigg[\epsilon_{\ast}(M_{\rm min})\,M_{\rm min} \\
    & + \int^{M_{\rm halo}}_{\rm M_{\rm min}} {\rm d}M^\prime_{\rm halo}\,f_{\rm b}\,\epsilon_{\ast}(M^\prime_{\rm halo})\,\mathcal{F}(\sigma_{\rm sf})\bigg], \nonumber
\end{align}
where $M_{\rm min}=10^8 \msun$ is the minimum halo mass we consider (the SFE below $M_{\rm min}$ plateaus at a constant value) that does not affect our results, $R$ is the mass return fraction of stars taken to be $0.1$ for the young and low-metallicity stellar populations at high redshifts~\citep[e.g.][]{Hopkins2018,Hopkins2023-fire3,Feldmann2024}, $\mathcal{F}(\sigma_{\rm sf})$ is the factor considering the difference between mean SFR and median SFR with the logarithm scatter $\sigma_{\rm sfr}$ in unit of dex $\mathcal{F}(\sigma_{\rm sfr}) = \exp{\left[(\ln{10}\,\sigma_{\rm sfr})^2/2\right]}$.
If the UV variability is dominated by the burstiness of star-formation, we have $\sigma_{\rm sfr} = \sigma_{\rm UV}/2.5 = 0.3\,{\rm dex}$. We have verified that this approach gives consistent stellar-to-halo mass ratios with abundance-matching results~\citep[e.g.][]{Behroozi2013,Behroozi2019,RP2017}. This stellar-to-halo mass relation will be used to derive the stellar mass function of galaxies in the empirical model only.

\section{Results}

As an overview of the simulations, we show in Figure~\ref{fig:img} dark matter surface density maps centered on the most massive halo in the simulation volume for both the $\Lambda$CDM and EDE cosmologies at $z\simeq 9$. Each map covers a $25\,{\rm cMpc}$ field of view with a projection thickness also $25\,{\rm cMpc}$. In Figure~\ref{fig:img2}, we show the difference in dark matter surface density, highlighting enhanced matter clustering in EDE. Structures around the knots of the cosmic web are up to $\sim 50\%$ denser in EDE compared to $\Lambda$CDM. Overlaid on the maps are the positions of moderately bright (intrinsic $M_{\rm UV} \leq -18$ mag) and very bright (intrinsic $M_{\rm UV} \leq -20$ mag) galaxies. As a result of more rapid early structure formation, EDE yields a higher abundance of bright galaxies, which contributes to the enhanced UV luminosity and stellar mass functions (see Section~\ref{sec:uvlf} and \ref{sec:smf} below). Interestingly, despite stronger underlying matter clustering in EDE, the bright galaxies appear less clustered. This counterintuitive trend arises from the reduced galaxy/halo bias in EDE: galaxies of fixed UV luminosity reside in more abundant, less rare halos compared to $\Lambda$CDM, leading to a lower clustering power~\citep{Klypin2021}.

\begin{figure*}
    \raggedright
    \includegraphics[width=0.49\linewidth, trim={0.1cm 0.3cm 0.3cm 0.3cm}]{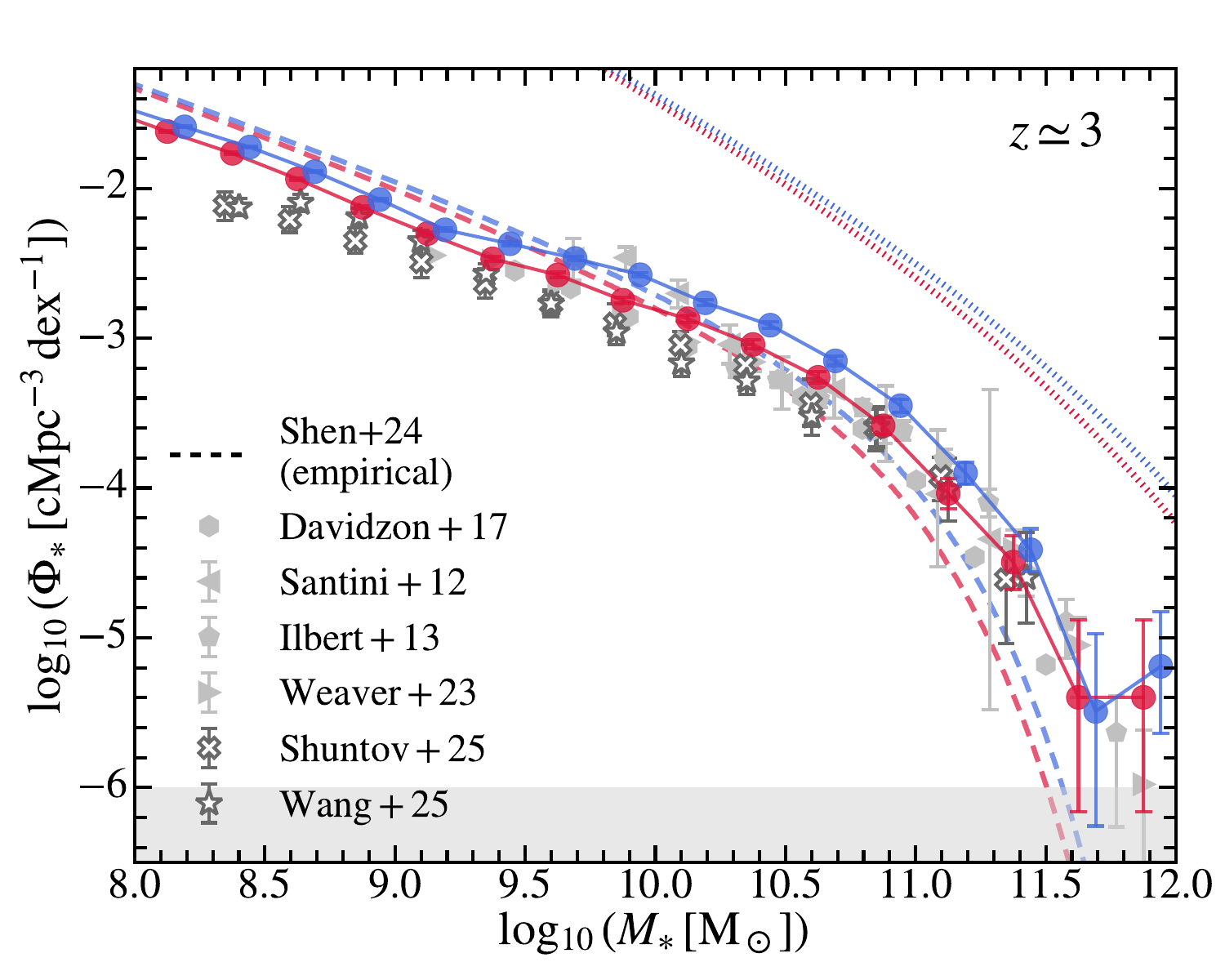}
    \includegraphics[width=0.49\linewidth, trim={0.1cm 0.3cm 0.3cm 0.3cm}]{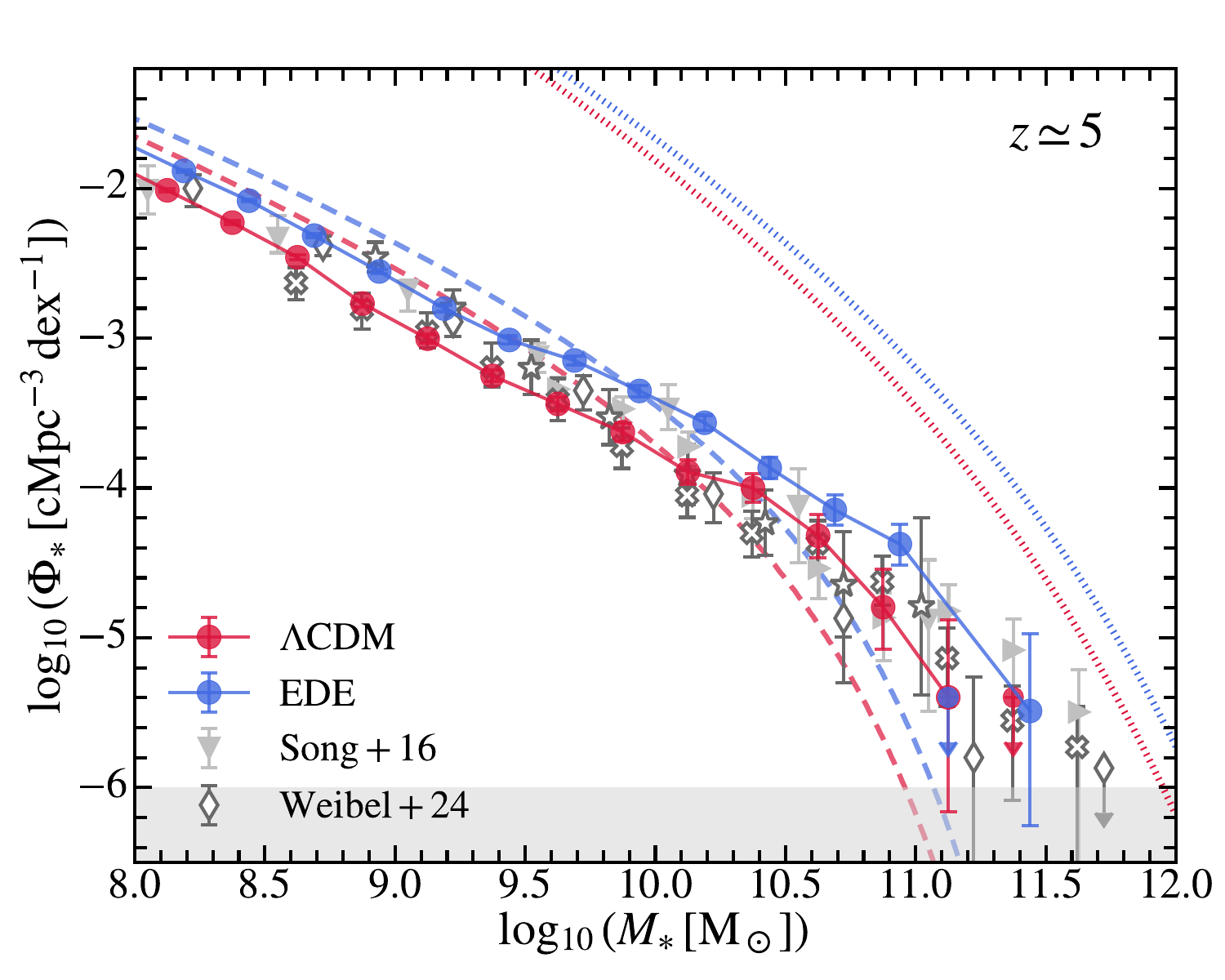}
    \includegraphics[width=0.49\linewidth, trim={0.1cm 0.3cm 0.3cm 0.3cm}]{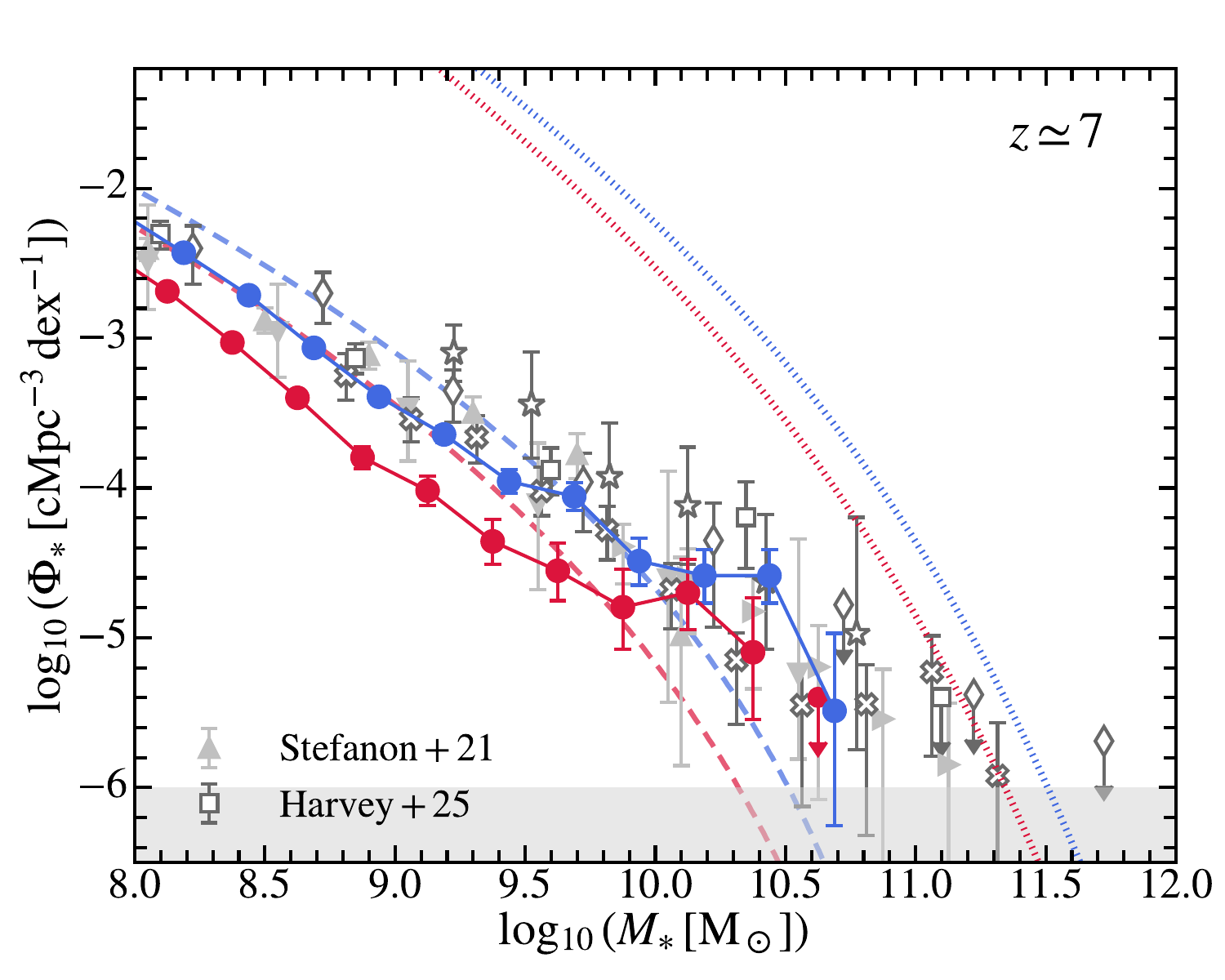}
    \includegraphics[width=0.49\linewidth, trim={0.1cm 0.3cm 0.3cm 0.3cm}]{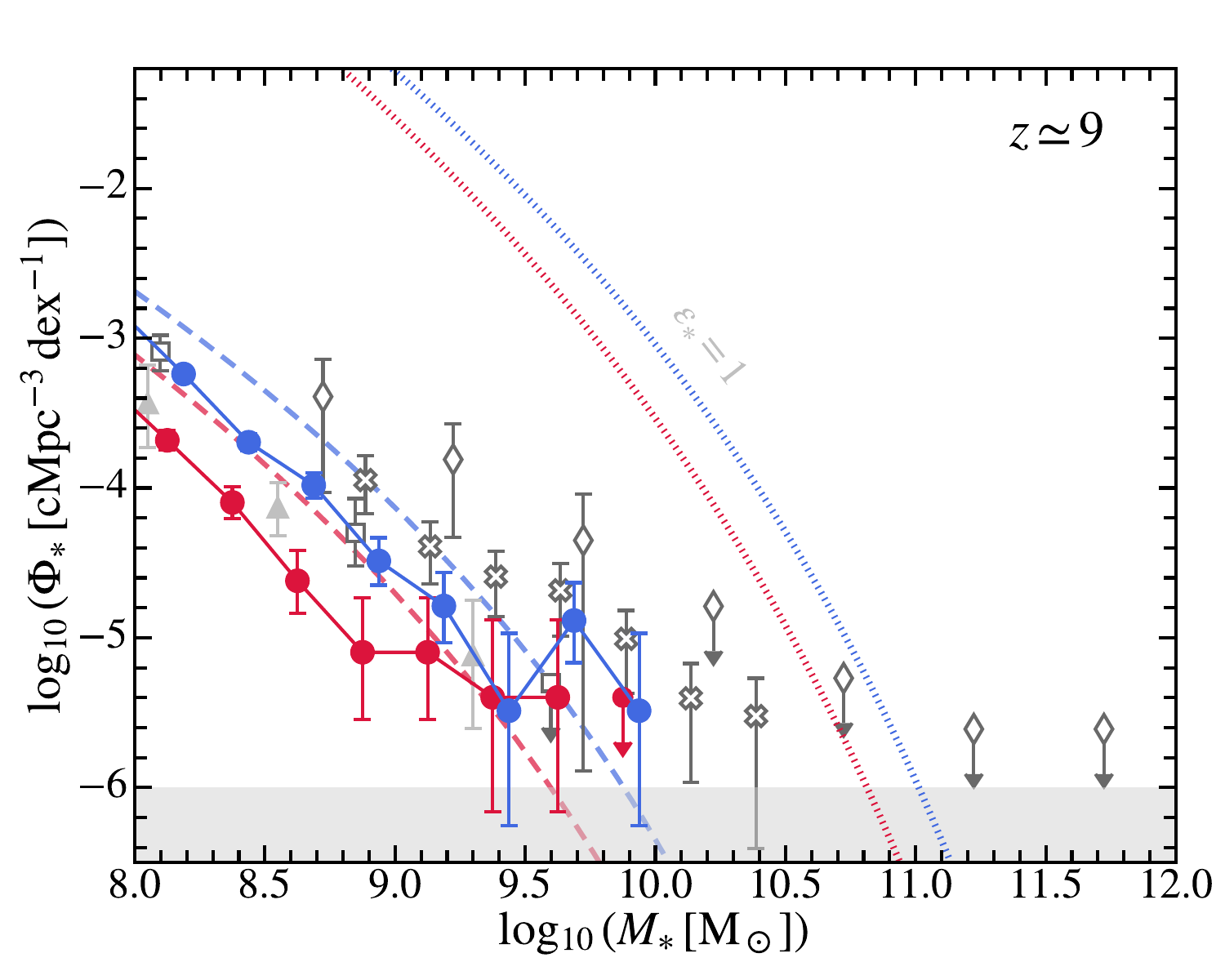}
    \caption{Galaxy stellar mass function at $z=3,5,7,$ and 9 in the $\Lambda$CDM and EDE runs. They are shown in solid circles, and Poisson errors are computed using the formula from \citet{Gehrels1986}. We compare them to the pre-JWST observational constraints from \citet{Santini2012,Ilbert2013,Song2016,Davidzon2017,Stefanon2021,Weaver2023} (in solid gray points) and more recent JWST constraints from \citet{Weibel2024,WangT2024,Harvey2025,Shuntov2025} (in open black points). Predictions from the empirical model~\citep{Shen2023,Shen2024b} are shown in dashed lines along with the theory limit ($\epsilon_{\ast}=1$) in dotted lines for reference. At $z\simeq 3$, the EDE run almost converges to the $\Lambda$CDM results, and both of them achieve decent agreement with observations. At higher redshifts, the EDE run produces enhanced galaxy abundance at all stellar masses, and the differences to $\Lambda$CDM are close to what we predict using the simple empirical model. Notably, at $z\simeq 7$ and 9, the EDE model predictions show reasonable agreement with observational constraints, in particular the ones obtained by JWST.
    }
    \label{fig:smf}
\end{figure*}

\subsection{UV luminosity functions}
\label{sec:uvlf}

We first examine the widely discussed excess of UV-bright galaxies at $z\gtrsim 10$ reported by JWST. In Figure~\ref{fig:uvlf}, we show the rest-frame UV luminosity functions of galaxies in the $\Lambda$CDM and EDE simulations from $z\simeq 4$ to $14$. Predictions are truncated at $M_{\rm UV} \simeq -17$, above which galaxies are not well resolved. We show the binned estimations along with $1\sigma$ Poisson error bars computed following \citet{Gehrels1986}. They are compared to the observational constraints compiled in \citet{Shen2024b}, which includes pre-JWST observations~\citep{McLeod2016,Oesch2018,Morishita2018,Stefanon2019,Bowler2020,Bouwens2021}, JWST photometry-based constraints~\citep{Castellano2022, Finkelstein2022, Naidu2022, Bouwens2023b, Bouwens2023a, Donnan2023, Harikane2023, Leetho2023, Morishita2023, Perez2023, Adams2023b, Robertson2024, McLeod2024, Donnan2024, Casey2024}, and the constraints based only on spectroscopically-confirmed galaxies in \citet{Harikane2024-spec,Harikane2024b-spec}. In addition to this compilation, we add the recent measurement at $z\simeq 14.5$ based on the spectroscopically-confirmed galaxy MoM-14~\citep{Naidu2025}. 

For predictions in the EDE model, the ``interpreted'' galaxy luminosities and number densities by an observer assuming $\Lambda{\rm CDM}$ should have additional corrections as
\begin{align}
    & \Phi^{\prime} = \Phi  \times  \dfrac{({\rm d}V/{\rm d}z)_{\rm EDE}}{({\rm d}V/{\rm d}z)_{\Lambda{\rm CDM}}}, \nonumber \\
    & M_{\rm UV}^{\prime} = M_{\rm UV} + 2.5 \log_{10}{\left( (D^{\rm EDE}_{\rm L} / D^{\Lambda{\rm CDM}}_{\rm L})^{2}  \right)},
\end{align}
where values with prime denote the measured values by the observer, $({\rm d}V/{\rm d}z)$ and $D_{\rm L}$ are the differential comoving volume and luminosity distance at the redshift of interest. Galaxy stellar mass derived from SED fitting will have the same correction factor as the luminosities. In the rest of the paper, whenever we make comparisons with observations, galaxy stellar masses, luminosities, number or mass densities in the EDE cosmology will all be corrected.

As shown in Figure~\ref{fig:uvlf}, the UV luminosity functions are systematically higher in EDE compared to $\Lambda$CDM. The enhancement of galaxy number density is from $\simeq 0.2$ dex at $z\simeq 4$ to $\simeq 0.8$ dex at $z\simeq 14$. For comparison, we overlay the predictions from the empirical model. The amplitude of the enhancement in UV luminosity functions broadly agrees with empirical model predictions. However, small differences remain likely due to the small differences in the assumed SFE and the younger stellar ages at higher redshifts (which affects $\kappa_{\rm UV}$).  

At $z\simeq 4$, the difference between the EDE and $\Lambda$CDM predictions has already shrunk to the level of the observational data scatter (plus the uncertainties in dust attenuation as discussed in Section~\ref{sec:empirical}), with the EDE model providing slightly better agreement with observations. This gap is expected to vanish at even lower redshifts. The results at $z\simeq 5$ and $6$ are qualitatively similar to the $z\simeq 4$ case. Towards cosmic dawn, at $z\simeq 9,12$, and 14, the UV luminosity functions predicted by our $\Lambda$CDM simulation start to fall significantly below the observational constraints, by up to $\sim 1$ dex at some luminosities. These results are consistent with other simulations using the IllustrisTNG model in the $\Lambda$CDM cosmology~\citep[e.g.][]{Kannan2022-thesan,Kannan2023,Pakmor2023}. In contrast, the UV luminosity functions from our EDE simulation are more consistent with observations across the entire redshift range. The large observed abundance of UV-bright galaxies at $z\gtrsim 12$ is successfully reproduced. Notably, no fine-tuning has been introduced: all the EDE model parameters were fixed by fitting the CMB data~\citep{Smith2022}, and the galaxy formation model is the out-of-the-box IllustrisTNG model calibrated at low redshifts. Together, these results indicate that EDE offers a simple and effective explanation for the excess of UV-bright galaxies at cosmic dawn, while naturally converging to $\Lambda$CDM predictions at later times.

\subsection{Stellar mass functions and massive galaxy outliers}
\label{sec:smf}

\begin{figure}
    \centering
    \includegraphics[width=\linewidth, trim={0 1cm 0 0}]{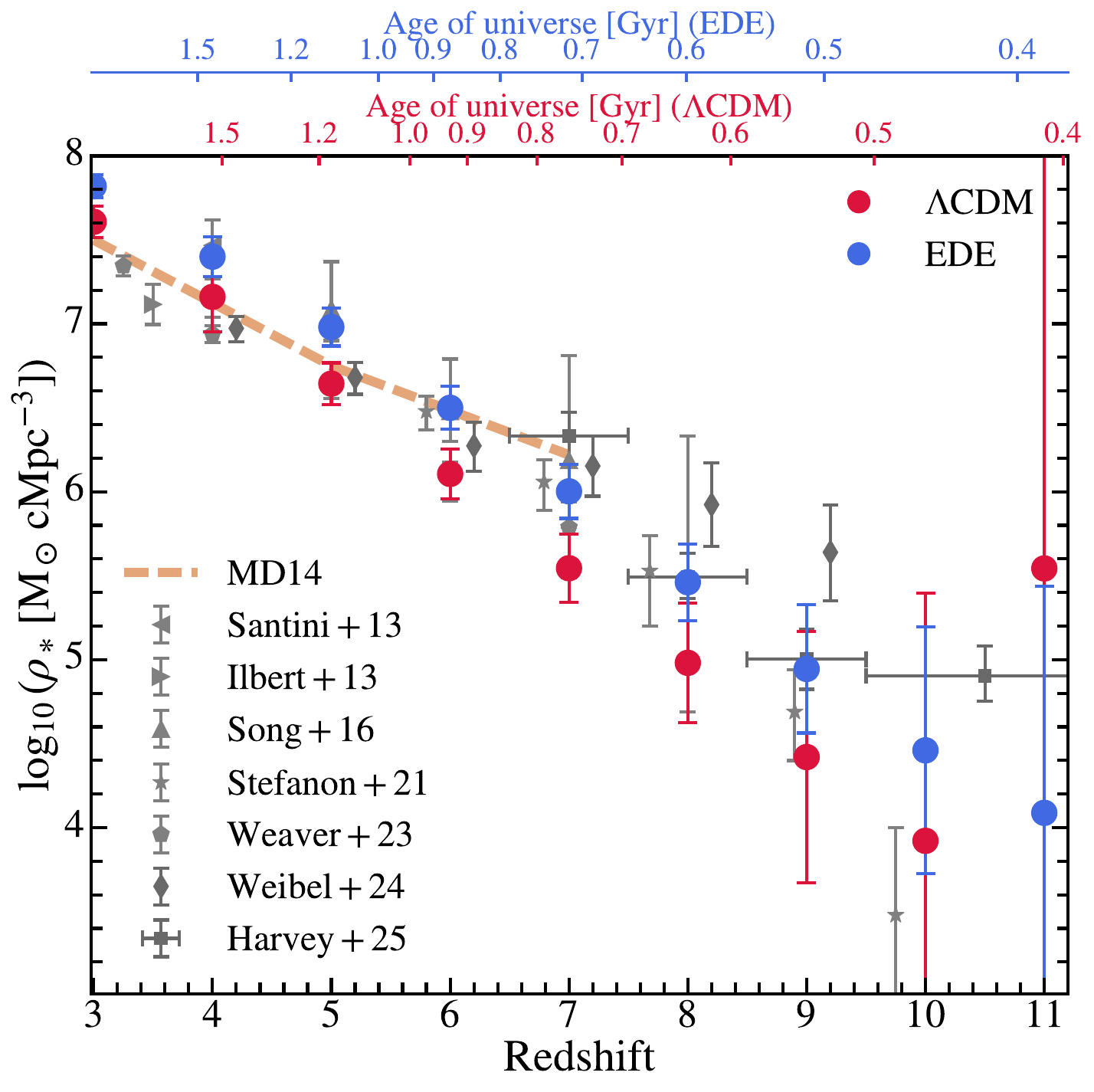}
    \caption{Cosmic stellar mass density versus redshift. We integrate the Schechter function fitted at each redshift with $M_{\rm min}=10^{8}\msun$ and $M_{\rm max}=10^{13}\msun$ to obtain the cosmic stellar mass density. The top two axes show the age of the universe in the two cosmologies. Observational constraints from \citet{Santini2012,Ilbert2013,Song2016,Stefanon2021,Weaver2023,Weibel2024,Harvey2025} are shown for comparison. The orange dashed line shows the stellar mass density in \citet{Madau2014} derived from integration of cosmic SFR density. Compared to $\Lambda$CDM, the EDE run predicts enhanced stellar mass densities at progressively higher redshifts, which agrees better with observations at $z\gtrsim 6$. The difference between the two models diminishes ($\lesssim 0.2$ dex) at $z\simeq 3$.}
    \label{fig:smd_vs_z}
\end{figure}

\begin{figure}
    \centering
    \includegraphics[width=\linewidth]{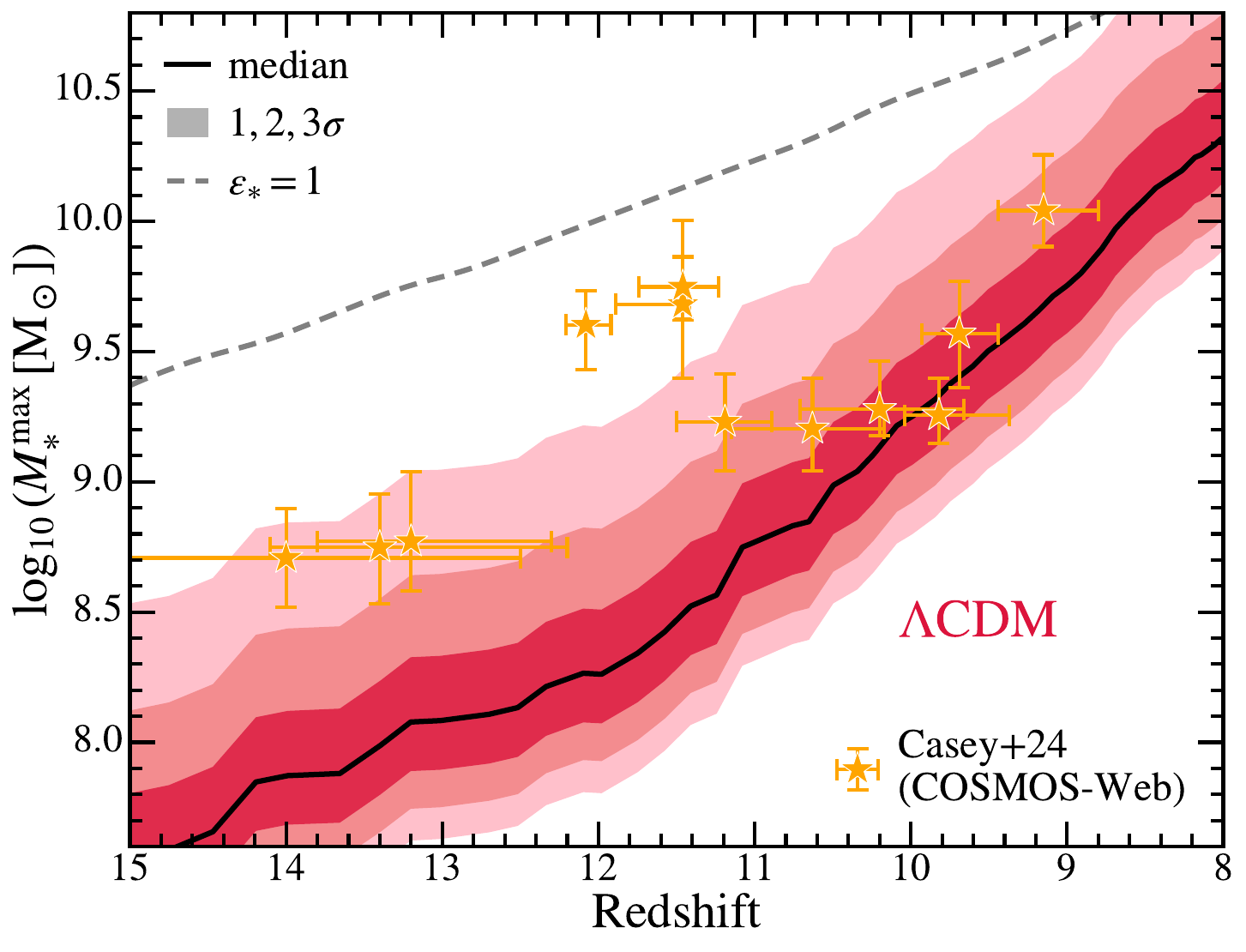}
    \includegraphics[width=\linewidth]{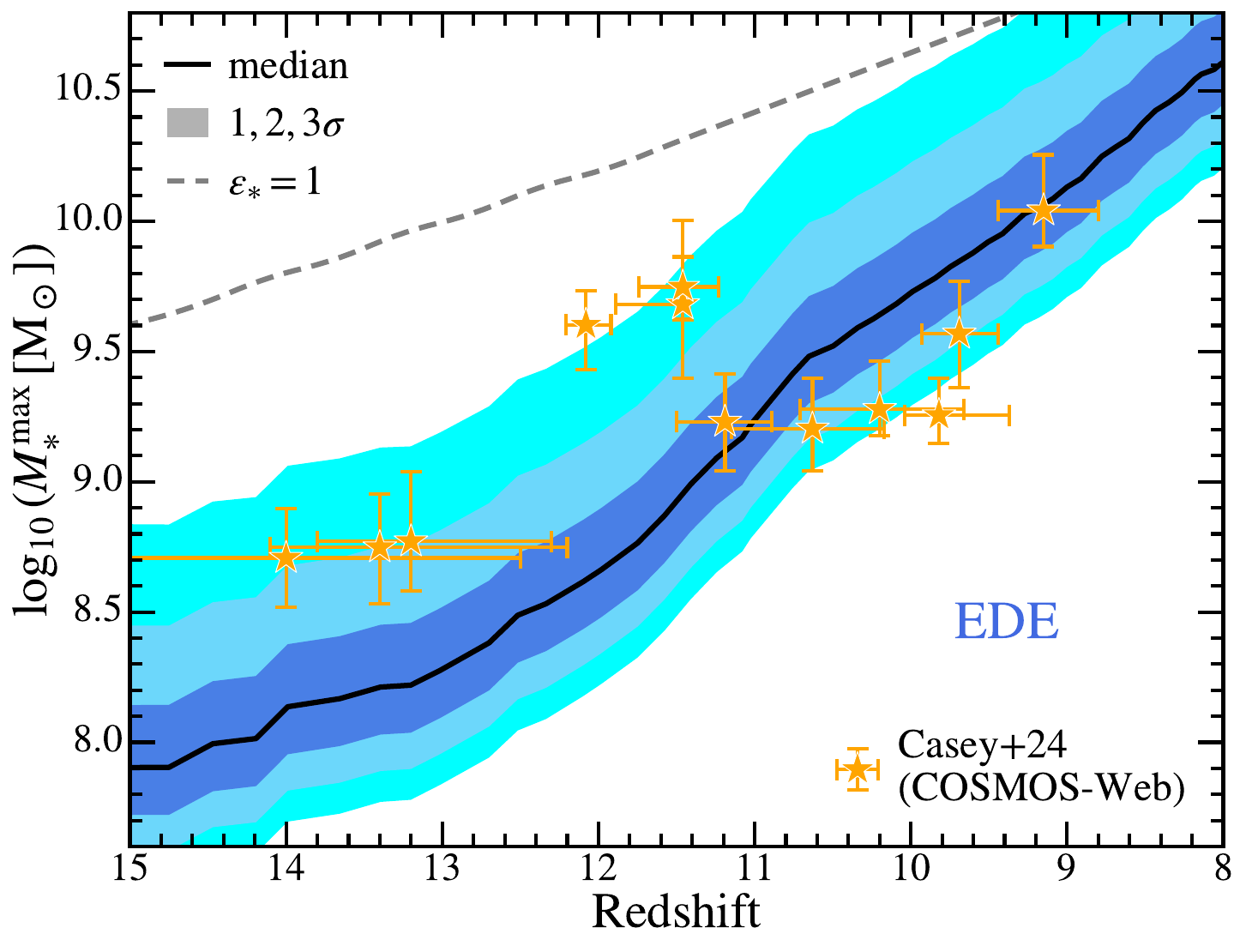}
    \caption{Maximum stellar mass of galaxies within a volume probed by the COSMOS-Web survey~\citep[$\simeq 0.28\,{\rm deg}^2$ used in ][]{Casey2024}. The solid lines in the top and bottom panels show the estimates based on our $\Lambda$CDM and EDE simulations, respectively. The shaded regions display $1,2,3\sigma$ dispersion computed using the extreme value statistics~\citep{Lovell2023} and the dashed lines show the theory limit, which is the maximum baryon mass in a halo in the volume. We compare the results to the observed bright galaxy sample in \citet{Casey2024} shown in yellow stars. Most of the observed samples at $z\lesssim 11$ are $1-2\sigma$ above the $\Lambda$CDM prediction, with extreme ones at $z\simeq 11-12$ falling outside the upper $3\sigma$ contour. In EDE, all of the samples are within $3\sigma$.}
    \label{fig:mmax}
\end{figure}

Another widely discussed tension raised by JWST observations is the abundance of massive galaxies. We begin by examining the galaxy stellar mass functions. Figure~\ref{fig:smf} shows the stellar mass functions from the $\Lambda$CDM and EDE runs at redshifts $z=3, 5, 7,$ and $9$. We compare these predictions against two sets of observational constraints: pre-JWST results from \citet{Santini2012,Ilbert2013, Song2016,Davidzon2017,Stefanon2021,Weaver2023}, and more recent JWST-based measurements from \citet{Weibel2024,WangT2024,Harvey2025,Shuntov2025}. While both sets of observations generally agree at lower redshifts and toward the low-mass end, JWST reveals a higher abundance of massive galaxies at $z \gtrsim 7$ with its better rest-frame optical coverage~\citep[e.g.][]{Weibel2024}. Contamination of faint obscured AGN, known as 
Little Red Dots (LRDs), can become relevant at $z\sim 6-8$ and $M_{\ast}\gtrsim 10^{10}\msun$~\citep[e.g.][]{Weibel2024,WangT2024,Harvey2025}, but most of the JWST constraints compiled here have already done some filtering of potential LRD contamination.

The empirical model again reproduces the relative offset between the two cosmologies, although it overpredicts (underpredicts) the abundance compared to the simulations at the low- (high-) mass end. At $z\simeq 3$, the EDE and $\Lambda$CDM predictions differ by $\lesssim 0.2$ dex overall and agree with observations well at the massive end ($M_{\ast}\gtrsim 10^{11}\msun$). A mild excess in the EDE model appears at $M_{\ast}\simeq 10^{10.5}\msun$, but this is comparable to the formal uncertainties in stellar mass estimates in SED fitting ($\gtrsim 0.2$ dex in $M_{\ast}$, e.g. \citealt{Wang2024a}) and uncertainties from mass definitions in theoretical works~\citep[e.g.][]{Pillepich2018b}. At higher redshifts, the difference between EDE and $\Lambda$CDM becomes more pronounced. The $\Lambda$CDM run underpredicts galaxy stellar mass functions by up to one dex at $z\sim 7-9$. The EDE run predictions show better agreement with observations at similar redshifts. The only observation that even the EDE model fails to match at roughly the $1\sigma$ level is the measurement at $z\simeq 9$ from \citet{Weibel2024}. It is also worth noting that even the most stringent observational limits on the stellar mass function do not require the maximal (unity) SFE in our empirical model. The comparison here shows the potential of EDE as a coherent solution to the excess of massive galaxies and UV-bright galaxies across a wide redshift range.

Another quantity commonly used to characterize the global evolution of the galaxy population is the cosmic stellar mass density $\rho_{\ast}$, defined as 
\begin{equation}
    \rho_{\ast}(z) \equiv \int_{M_{\rm min}}^{M_{\rm max}} \Phi_{z}(M_{\ast})\,M_{\ast}\,{\rm d}\log{M_{\ast}},
\end{equation}
where $\Phi_{z}$ denotes the stellar mass function at redshift $z$. The integration limits, $M_{\rm min}$ and $M_{\rm max}$, are set to $10^8\,\msun$ and $10^{13}\,\msun$, respectively, following conventions in the observational literature~\citep[e.g.][]{Song2016,Stefanon2021,Weibel2024}. We fit the stellar mass function at each redshift (above $10^8\,\msun$ to avoid the low-mass cutoff due to the resolution limit) with a Schechter function and then compute the integrated stellar mass density. At $z<5$, the break stellar masses can be successfully obtained through fits with standard errors $<1$ dex. At $z\geq 5$, we fix the break stellar mass of the Schechter function to $10^{11}\msun$, which is consistent with observational results~\citep[e.g.][]{Shuntov2025}. The exact value of the break stellar mass has a relatively minor impact on the integrated stellar mass density at these redshifts, as low-mass galaxies dominate the total mass budget.

In Figure~\ref{fig:smd_vs_z}, we show the cosmic stellar mass density as a function of redshift for the $\Lambda$CDM and EDE runs, compared against observational constraints from \citet{Santini2012,Ilbert2013,Song2016,Stefanon2021,Weaver2023,Weibel2024,Harvey2025}, again labelled with different colors to distinguish pre- and post-JWST measurements. Consistent with our findings for the stellar mass functions, the stellar mass density is systematically higher in the EDE simulations than in $\Lambda$CDM, with a difference of $\simeq 0.6$ dex at $z\simeq 10$ that narrows to $\simeq 0.2$ dex by $z\simeq 4$. The EDE model shows better agreement with observational constraints at $z \gtrsim 6$. In this figure, we also include the age of the universe as a function of redshift in the two cosmologies (see \citealt{BK2021}). The difference in lookback time between the models is generally less than $50$ Myr at $z \gtrsim 6$, which is smaller than the typical e-folding growth time of star-forming galaxies on the main sequence at these redshifts~\citep[e.g.][]{Speagle2014,Popesso2023}. This reinforces that the enhanced galaxy/halo abundance in EDE is primarily driven by the amplified small-scale power spectrum and the earlier formation of low-mass haloes. It is not a consequence of having more time to form structures, and in fact, EDE reduces the age of the Universe at all redshifts relative to $\Lambda$CDM.

Thus far, we have focused on the overall and integrated properties of galaxies. We now turn to evaluating the likelihood of detecting individual massive galaxy outliers in observational surveys. As a case study, we consider the COSMOS-Web field ($\simeq 0.28\,\mathrm{deg}^2$, \citealt{Casey2024}). Assuming a redshift slice of $\Delta z = 2$, our $(100\,\mathrm{cMpc})^3$ simulation volume corresponds to a field of view of $\sim 0.07\,\mathrm{deg}^2$ at $z \simeq 8$ and $\simeq 0.10\,\mathrm{deg}^2$ at $z \simeq 12$ in the $\Lambda$CDM cosmology. In the EDE cosmology, the corresponding values are $\simeq 0.08\,\mathrm{deg}^2$ at $z \simeq 8$ and $\simeq 0.12\,\mathrm{deg}^2$ at $z \simeq 12$. We first measure the stellar mass of the most massive galaxy in our simulation volume. To estimate the most massive galaxy expected in a given observed field, we apply a simple translation using our empirical model
\begin{equation}
\dfrac{M_{\ast, {\rm max}}^{\rm obs}}{M_{\ast,{\rm max}}^{\rm sim}} = \dfrac{M_{\ast}\left({\rm CDF}(>M_{\ast}) = 1/V_{\rm obs}\right)}{M_{\ast}\left({\rm CDF}(>M_{\ast}) = 1/V_{\rm sim}\right)},
\end{equation}
where ${\rm CDF}(>M_{\ast})$ is the cumulative stellar mass function computed from our empirical model, which can go beyond the mass range directly covered in the simulations, $V_{\rm sim}$ and $V_{\rm obs}$ are the comoving volume covered by the simulation and the observational survey. We further compute the uncertainties of $M^{\rm obs}_{\ast\,{\rm max}}$ using the extreme value statistics as in \citet{Lovell2023}. In Figure~\ref{fig:mmax}, we compare the expected maximum stellar mass in the COSMOS-Web field estimated based on our simulations to the observed massive galaxy candidates reported by \citet{Casey2024}. The shaded regions of different colors show $1,2,3\sigma$ uncertainties. The observed massive galaxies at $8 \lesssim z \lesssim 11$ are consistent with the median $M_{\ast\,{\rm max}}$ in EDE. The most massive candidates at $z\gtrsim 11$ are still consistent with the EDE model prediction at a $2-3\sigma$ level. This is encouraging as these galaxies are selected to be the brightest ones in the COSMOS-Web field. On the contrary, the $\Lambda$CDM prediction will be off from the observed galaxies at $z\simeq 11-12$ at more than a $3\sigma$ level.

\begin{figure}
    \centering
    \includegraphics[width=\linewidth, trim={0 1cm 0 0}]{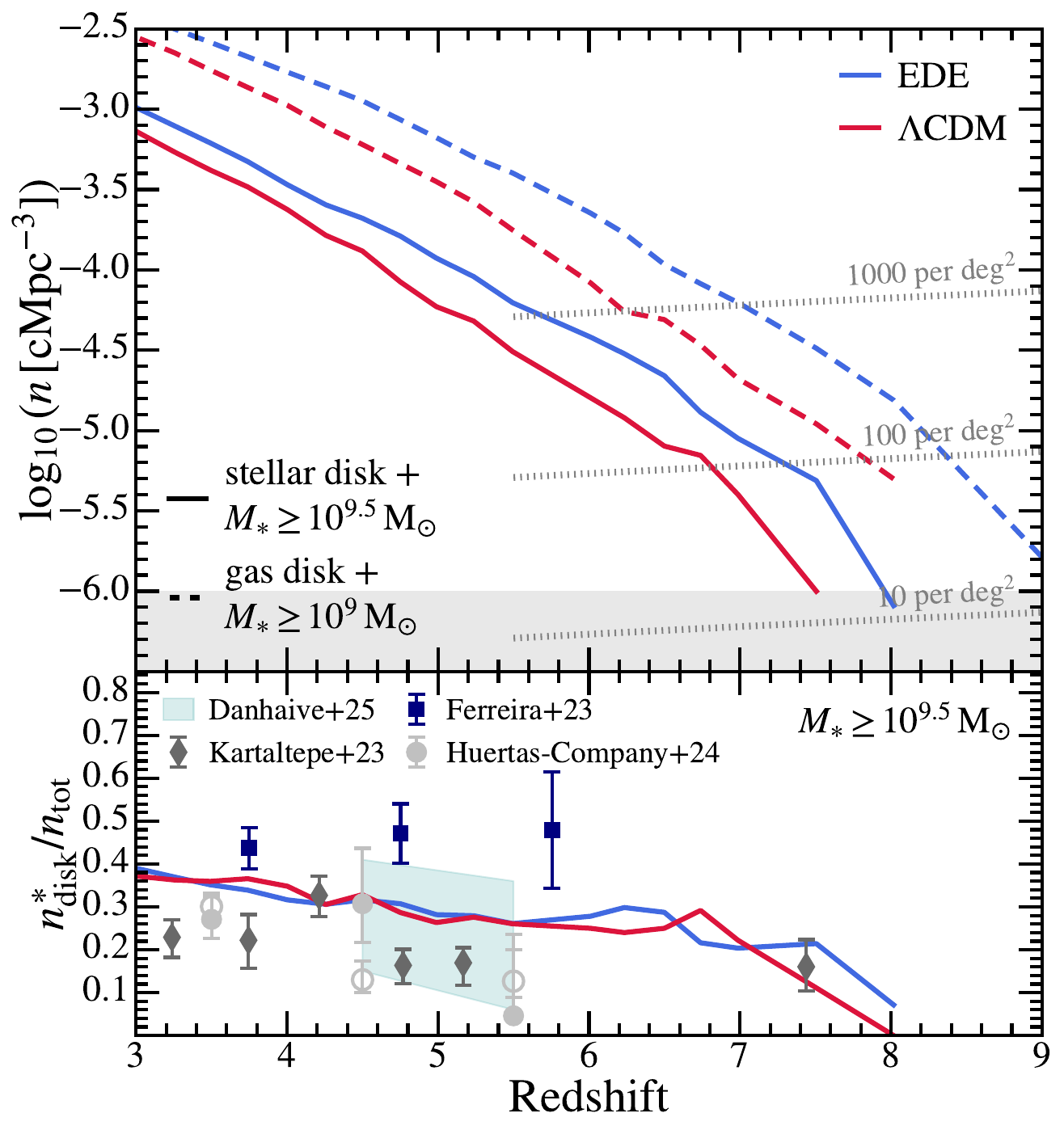}
    \caption{{\it Top}: Number density of disky galaxies in the universe versus redshift in the $\Lambda$CDM and EDE runs. We show the number densities of massive galaxies ($M_{\ast}\geq 10^{9.5}\msun$) with stellar disks in solid lines, and those of galaxies ($M_{\ast}\geq 10^{9}\msun$) with gaseous disks in dashed lines. Disky galaxies in EDE appear systematically earlier, and the number density of disky galaxies is enhanced by roughly half a dex at $z\gtrsim 6$. Dotted lines indicate the number density that corresponds to $10-1000$ objects per ${\rm deg}^2$ with a survey depth of $\Delta z=1$. {\it Bottom}: Fraction of disky galaxies among all galaxies with $M_{\ast} \geq 10^{9.5}\msun$ versus redshift. We compare it with observational constraints from \citet{Ferreira2023,Kartaltepe2023,Huertas-Company2024,Danhaive2025}, of which the mass completeness limits roughly match our mass cut. The simulation results reproduce the trends with a fairly stable disky fraction from $\sim 40\%$ at $z\simeq 3$ to $\sim 20\%$ at $z\simeq 7-8$ when the first disky galaxy appears in our simulation volume. The $\Lambda$CDM and EDE runs do not show dramatic differences, reflecting the strong correlation of galaxy morphology with stellar mass in the IllustrisTNG model.}
    \label{fig:disk}
\end{figure}

\subsection{Early emergence of disky galaxies}
\label{sec:disk}

The luminosity and stellar mass function analyses shown in previous sections have demonstrated the more rapid structure formation in EDE. One natural deduction is that the morphological transition of galaxies from low-mass, dispersion-supported systems to massive, disky galaxies will also arrive earlier~\citep[e.g.][]{El-Badry2018,Tacchella2019}. This would be, at least qualitatively, in line with the large fraction of disk-like sources in JWST imaging data~\citep[e.g.][]{Robertson2023b}, the discovery of early dynamically cold molecular gas disks by ALMA~\citep[e.g.][]{Rizzo2021,Parlanti2023,Rowland2024,Fujimoto2024}, and similar findings using ionized gas~\citep[e.g.][]{Nelson2024,Xu2024,Danhaive2025}.

To investigate this scenario, we perform characterization of disky galaxies based on angular momentum criteria as introduced in Section~\ref{sec:sim}. In Figure~\ref{fig:disk}, we show the number density of disky galaxies identified in the $\Lambda$CDM and EDE simulations as a function of redshift. We present galaxies with stellar disks above $10^{9.5}\msun$ and those with gaseous disks above $10^{9}\msun$. These mass thresholds are chosen to match roughly the stellar mass range probed in observations (as will be discussed below) and do not affect our qualitative conclusions. In the bottom subpanel, we show the fraction of disky galaxies at $M_{\ast}\geq 10^{9.5}\msun$ versus redshift. In both cosmologies, stellar disks in massive and mature galaxies begin to emerge very early around $z\simeq 7-8$ in the $(100\,{\rm cMpc})^{3}$ volume, but their abundance increases more rapidly in the EDE simulation. At $z\simeq 6-7$, the number density of disky galaxies in EDE exceeds that in $\Lambda$CDM by half a dex, reflecting the accelerated buildup of rotationally supported stellar structures in the earlier-forming halos. The rapid assembly of massive halos and their associated cold gas reservoirs in EDE facilitates the formation of stable disks at earlier times. This is visible from the number density of galaxies with gas disks, which is also systematically enhanced in the EDE run with roughly the same magnitude. The elevated abundance of disky galaxies in EDE primarily reflects the increased number of massive galaxies, while the correlation between diskiness and stellar mass remains unchanged (see Figure~\ref{fig:fdisk_z6} in the Appendix). Therefore, the disky galaxy fraction is not sensitive to the choice of cosmology and instead is a prediction of the IllustrisTNG galaxy formation model.

We compare our results to the recent morphological studies based on JWST imaging data in \citet{Kartaltepe2023,Ferreira2023,Huertas-Company2024}. \citet{Ferreira2023} and \citet{Kartaltepe2023} visually classified the morphology of galaxies and reported the disk galaxy fractions at $M_{\ast}\gtrsim 10^{9}\msun$ based on the CEERS survey data that overlapped with the CANDELS data. The mass-completeness limit of the overlapping CANDELS data is at least $10^{9.5}\msun$~\citep{Duncan2019}, which is roughly consistent with the stellar mass cut we adopt. The ``visual'' disk criterion in \citet{Ferreira2023} is likely less stringent than we used and results in overall higher disk fraction at $z\simeq 4-6$. For example, they compared results to the EAGLE and FLARES simulations, and the disk criterion used was a spheroid-to-total mass ratio ($\sim 1 - \,$D/T) less than $0.75$. For \citet{Kartaltepe2023}, we take their ``disk-only'' fraction for comparison. \citet{Huertas-Company2024} used convolution neural networks to classify galaxy morphology in the CEERS survey, and we take their results in the two $M_{\ast}\geq 10^{9.8}\msun$ bins for comparison. Despite slightly different definitions of disks among observational studies and compared to simulations, we reproduce a large and consistent disk fraction of $\gtrsim 20\%$ out to $z\simeq 7$ ($z\simeq 7.5$) in the $\Lambda$CDM and EDE runs. It increases to about $30-40\%$ at $z\simeq 3-4$, which agrees with the observational constraints.

In addition to stellar disks identified based on imaging data, several works have reported disky systems based on ionized gas kinematics~\citep[e.g.][]{Nelson2024,Xu2024,Danhaive2025}. For example, \citet{Danhaive2025} measured morphological and kinematic properties of ionized gas in galaxies as probed by H$\alpha$. They reported a substantial but not dominant fraction of rotationally supported galaxies at $z\sim 4-6$ in the stellar mass range $10^{9}-10^{10}\msun$. We overlay their constraints in the bottom panel of Figure~\ref{fig:disk}, which are consistent with our simulation predictions. 

The $[{\rm C}\,{\rm II}]$ line accessible through ALMA also enables the detection of dynamically cold molecular disks at high redshifts. A remarkable example is REBELS-25~\citep{Rowland2024} at $z\simeq 7.3$ with $M_{\ast} = 8_{-2}^{+4}\times 10^{9}\msun$, which is one of the $\sim 40$ bright galaxies observed by the ALMA-REBELS program in a $2+5\,{\rm deg}^2$ wide field~\citep{Bouwens2022}. REBELS-25 has a high ratio of ordered-to-random motion, $V_{\rm max}/\sigma = 11_{-5}^{+6}$, where $V_{\rm max}$ is the maximum rotation velocity and $\sigma$ is the mean velocity dispersion of ISM gas. Another example is the five rotating galaxies at $z\simeq 4-5$ reported in \citet{Roman-Oliveira2023} selected from public datasets with $[{\rm C}\,{\rm II}]$ line emission in the ALMA archive. The SFRs of these galaxies ($\sim 200-5000\msun/{\rm yr}$) are comparable to the most massive galaxies in our simulation box. They have $V_{\rm max}/\sigma$ values $\sim 3-10$. A direct comparison of the number density of this type of source is hard due to the uncertainties in the selection function in observations. These galaxies are the brightest galaxies that lie systematically above the main sequence~\citep{Rowland2024} and have large sizes with resolved $[{\rm C}\,{\rm II}]$ observation, which are likely hosted by high-spin haloes~\citep[e.g.][]{Mo1998,Shen2024}. From our simulations, if we take a survey depth of $\Delta z=1$, we predict that $\sim 10$ gaseous disk is expected in a $1\,{\rm deg}^2$ field already at $z\simeq 9$ in the EDE model, and this number increases rapidly to $\sim 1000$ at $z\simeq 7$. In $\Lambda$CDM, the number of these galaxies will be half a dex smaller. However, in both cases, the number densities are much larger than the disk galaxy samples revealed by ALMA observations so far.

One remaining puzzle is the dynamical coldness of these gas disks in observations. The velocity dispersions measured in the original IllustrisTNG simulation~\citep{Pillepich2019} have reached $\gtrsim 60\kms$ at $z\sim 5$ in the mass range of interest, as opposed to the low values ($\sigma \sim 30\kms$) in REBELS-25. However, the IllustrisTNG model does not resolve the cold molecular phase of the ISM, and the velocity dispersions are measured for the warm, ionized gas traced by H$\alpha$. It has been shown that the H$\alpha$-based velocity dispersion can be a factor of $2-3$ larger than the $[{\rm C}\,{\rm II}]$-based value~\citep[e.g.][]{Ubler2019,Fujimoto2024,Kohandel2024}. A direct comparison regarding this aspect will require future zoom-in simulations to resolve the multiphase structure of the ISM in these galaxies.

\subsection{Quenched galaxies and the SMBH population}
\label{sec:quenched}

\begin{figure}
    \centering
    \includegraphics[width=\linewidth, trim={0 1cm 0 0}]{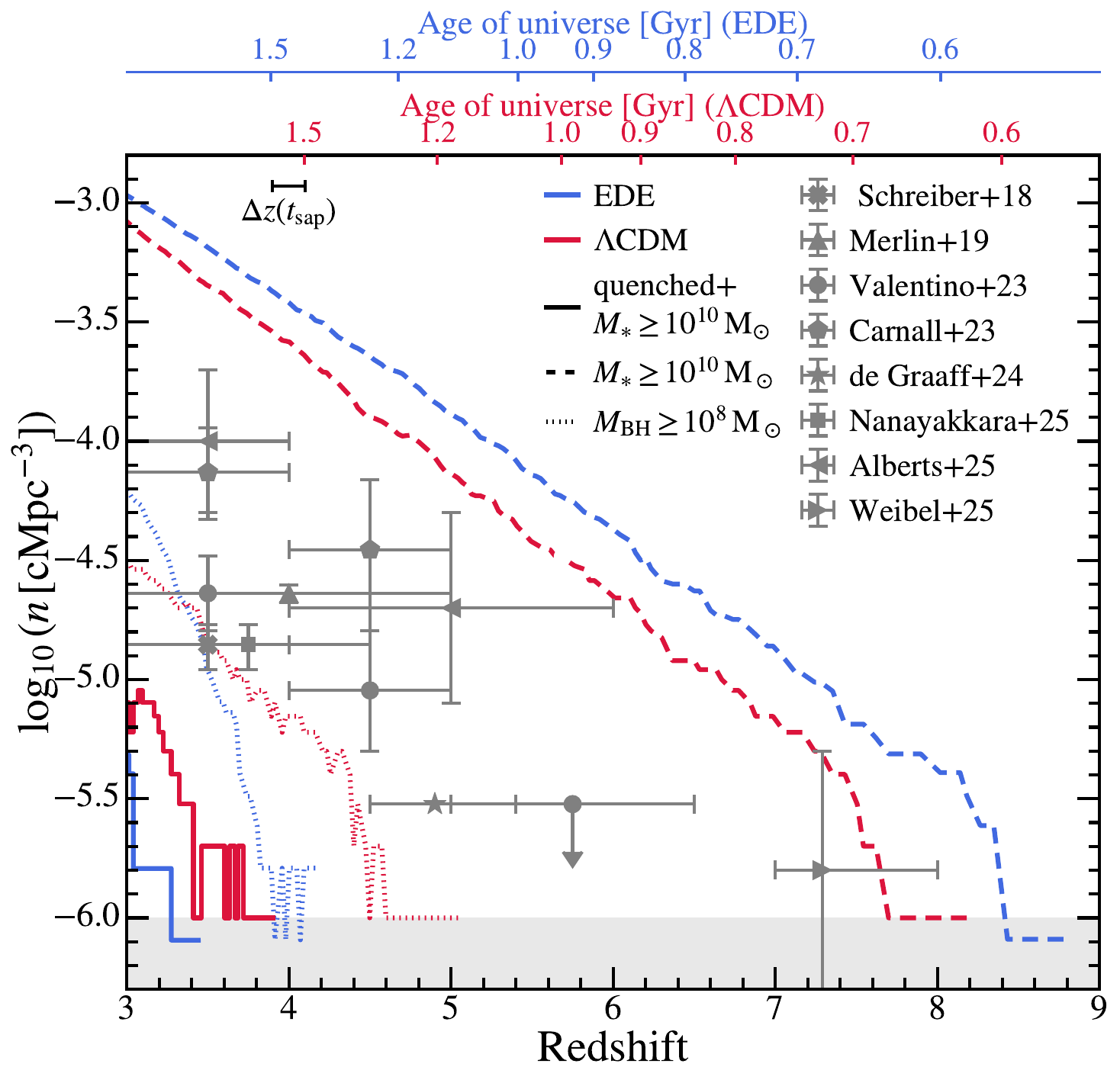}
    \caption{Number density of massive quenched galaxies versus redshift in the $\Lambda$CDM and EDE simulations. We show the number density of massive galaxies ($M_{\ast}\geq 10^{10}\msun$) in dashed lines, massive and quenched galaxies in solid lines, and massive SMBHs ($M_{\rm BH}\geq 10^{8}\msun$) in dotted lines. The quenched galaxy number density is compared to the observational estimates reported in \citet{Schreiber2018,Merlin2019,Carnall2023b,Carnall2023a,Valentino2023,Alberts2024,Nanayakkara2025,DeGraaff2025,Weibel2025}. The top two axes show the age of the universe in the two cosmologies. Although the number of massive galaxies is larger in the EDE model, quenched galaxies are rarer compared to $\Lambda$CDM. The number of quenched galaxies follows the emergence of massive SMBHs that are above the critical mass for quenching, which is a feature of the two-mode AGN feedback model in IllustrisTNG. The Salpeter time ($t_{\rm sap}$, e-folding mass growth time for Eddington-limit accretion) of SMBHs is short enough to be comparable to the difference in the age of the universe between the two cosmologies. Therefore, the most massive SMBHs grow more slowly in EDE due to the shorter total time for accretion (the age of the universe), which results in delayed quenching. We overlay a bar indicating the $\Delta z$ that corresponds to $t_{\rm sap}$ at $z=4$.}
    \label{fig:quenched}
\end{figure}

One counterintuitive result from our simulations is that quenched galaxies do not form earlier in the EDE model. Figure~\ref{fig:quenched} shows the redshift evolution of the number density of massive, quenched galaxies in the $\Lambda$CDM and EDE runs. We have restricted the selection to galaxies with $M_{\ast}\geq 10^{10}\msun$ and $M_{\rm halo}\geq 10^{11}\msun$ to avoid fake quenching in the low-mass end due to resolution effects. While massive galaxies with $M_{\ast} \gtrsim 10^{10}\msun$ appear earlier in EDE and maintain a consistently higher abundance than in $\Lambda$CDM down to $z\simeq 3$, quenched systems emerge at later times in EDE ($\simeq 3.5$ in EDE and $z\simeq 4$ in $\Lambda$CDM). 

In the IllustrisTNG model, quenching is strongly correlated to the growth and feedback of SMBHs~\citep[e.g.][]{Weinberger2018,Donnari2021a,KV2024}. Previous work has shown that quenching occurs once the cumulative feedback energy in kinetic mode surpasses the gravitational binding energy of the cold gas in the central region of a galaxy, which suppresses SMBH accretion and star-formation in the galaxy~\citep[e.g.][]{Weinberger2018,Terrazas2020,Zinger2020}. This is usually realized rapidly once the SMBH enters the kinetic mode. As implemented in IllustrisTNG, the kinetic mode is triggered when the Eddington ratio drops below $\lambda_{\rm kin} = \mathrm{MIN}\left[0.002\left(M_{\rm BH}/10^{8} \msun\right)^{2}, 0.1\right]$~\citep{Weinberger2017}. Due to the strong positive correlation of $\lambda_{\rm kin}$ with $M_{\rm BH}$, the kinetic mode gets turned on typically when $M_{\rm BH}$ exceeds a few times $10^{8}\msun$. Because the transition from thermal to kinetic mode is rapid, the onset of quenching is especially sensitive to the epoch at which the SMBH reaches its critical mass.

As shown in Figure~\ref{fig:mdot_and_phibh}, the most massive SMBHs at high redshifts in our simulations grow at the Eddington limit, with an e-folding growth timescale (the Salpeter time)
\begin{equation}
t_{\rm sap} = \dfrac{\sigma_{\rm T}\,c}{4\pi\,G\,m_{\rm p}}\,\dfrac{\epsilon_{\rm r}}{1-\epsilon_{\rm r}} \simeq 112.6 \Myr,
\end{equation}
where $\epsilon_{\rm r} = 0.2$ in the IllustrisTNG model, $c$ is the speed of light, $\sigma_{\rm T}$ is the Thomson scattering cross-section, and $m_{\rm p}$ is the proton mass. This growth timescale is short enough to be comparable to the cosmic age differences between $\Lambda$CDM and EDE: $t_{\Lambda\mathrm{CDM}} - t_{\mathrm{EDE}} \simeq 129.6,92.6,70.4\Myr$ at $z\simeq 3,4,5$, respectively. As a result, the most massive SMBHs are highly sensitive to cosmology through differences in the total accretion time, unlike the galaxy mass growth case investigated in the previous section. In EDE, due to the shorter age of the universe, the redshift at which SMBHs start to cross the critical mass for quenching becomes lower, which results in the delayed appearance of quenched galaxies. By contrast, lower-mass SMBHs grow at the Bondi–Hoyle rate in the IllustrisTNG model~\citep{Weinberger2017}, which depends on SMBH mass, local gas density, and (thermal) sound speed. The BH accretion in this regime is regulated by the thermal-mode AGN feedback counteracting gas cooling, and is therefore more tightly coupled to host galaxy properties rather than cosmology. The difference between the massive and low-mass SMBH populations is clearly illustrated in Figure~\ref{fig:mdot_and_phibh} in the Appendix. At fixed redshift, the EDE model produces more intermediate-mass SMBHs than $\Lambda$CDM, but paradoxically a lower number of the most massive SMBHs. However, comparing the simulations at fixed cosmic age removes this discrepancy, suggesting that the timing for SMBH growth (not the underlying structure abundance) affects the emergence of the first quenched galaxies.

In Figure~\ref{fig:quenched}, we compare the simulation results to the observational estimates compiled in \citet{Chittenden2025}, including \citet{Carnall2023b,Carnall2023a,Valentino2023,Alberts2024,Nanayakkara2025,DeGraaff2025,Weibel2025}. We adopt their most conservative estimates where available. Both the $\Lambda$CDM and the EDE runs underpredict the abundance of massive quenched galaxies in the universe by at least an order of magnitude at $z\gtrsim 3$. The observations require roughly $10\%$ of galaxies at $M_{\ast}\gtrsim 10^{10}\msun$ being quenched. This discrepancy points to limitations of the SMBH model in IllustrisTNG or missing physics that are responsible for galaxy quenching at high redshifts. One possible solution is a smoother transition between the two modes of feedback~\citep[e.g.][]{Lovell2023-quenching,Kimmig2025} and a weaker impact on the entropy of the surrounding gas, with quenching aided by e.g. stellar feedback~\citep{Kimmig2025}. Another solution is boosted SMBH growth at earlier times, which could be realized by super-Eddington accretion and/or weaker thermal mode feedback, and will result in larger SMBH-to-stellar mass ratios. This would be aligned with the recently discovered faint obscured AGN population at $z\gtrsim 4$~\citep[LRDs; e.g.][]{Kocevski2023,Matthee2024,Greene2024,Akins2024}, which lies at face value systematically above the local SMBH-to-stellar mass relations~\citep[e.g.][]{Pacucci2023,Harikane2023-agn,Juodzbalis2024} and displays signatures of super-Eddington accretion~\citep[e.g.][]{Pacucci2024,Inayoshi2024b,Maiolino2025,Naidu2025-bhstar}.

\section{Discussions and conclusions}
\label{sec:conclusion}

In this work, we have carried out large-volume $(100\,{\rm cMpc})^3$ cosmological hydrodynamic simulations to study galaxy formation at $z\geq 3$. One simulation adopts the standard $\Lambda$CDM cosmology, while the other incorporates a modest perturbation to the pre-recombination expansion history, realized in the form of an EDE component. The EDE model considered here has been proposed as a potential solution to the Hubble tension. It also serves as a representative example among a broader class of beyond-$\Lambda$CDM scenarios that accelerate the pre-recombination expansion. A generic consequence of such models is the enhancement of the small-scale matter power spectrum, which in turn accelerates the formation of non-linear structures at high redshifts. The cosmological hydrodynamic simulations presented in this work are the first to explore this class of models. Both simulations utilize the IllustrisTNG galaxy formation model with identical subgrid physics and numerical parameters, enabling a detailed comparison of galaxy properties across the two cosmologies. We find that EDE leads to more efficient structure formation in the early universe, resolving several emerging tensions revealed by JWST observations without requiring any fine-tuning of the galaxy formation model. Our main results are summarized below.

\begin{itemize}
    \setlength{\itemindent}{10pt}     
    \setlength{\itemsep}{5pt}
    \item \textbf{Excellent match to observed galaxy abundance at high redshifts}: The EDE simulation perfectly reproduces the UV luminosity functions measured by JWST out to $z\simeq 14$ and the stellar mass functions to $z\simeq 9$, while the $\Lambda$CDM simulation underpredicts them at $\sim 0.5 - 1$ dex level. This difference reflects the more rapid structure formation in the early universe in EDE, mainly due to the enhanced small-scale matter power spectrum. The observed extremely massive galaxies can be reconciled with the canonical galaxy formation model in EDE.

    \item \textbf{Natural convergence to $\Lambda$CDM at late times}: At low redshifts, the EDE model predictions converge to $\Lambda$CDM with typical differences in UV luminosity/stellar mass functions $\lesssim 0.2$ dex, and the agreement with low-redshift pre-JWST observations is preserved. There is essentially no additional tuning of the galaxy formation model, so that all the empirical success of the IllustrisTNG model is unaffected by this early-epoch cosmological perturbation.

    \item \textbf{Earlier disky galaxy formation}: Both stellar and gaseous disks emerge earlier and are more prevalent in EDE, with number densities enhanced by half a dex at $z\gtrsim 5$. This earlier appearance of disky galaxies is primarily driven by the increased abundance of massive galaxies in EDE, while the fraction of disky systems (at fixed stellar mass) remains largely unchanged. The disky galaxy fractions obtained in both runs are consistent with current observational constraints based on JWST imaging data and ionized gas kinematics. Our simulations predict a significantly higher number density of galaxies hosting gaseous disks than the ones currently discovered by ALMA through cold molecular gas tracers, suggesting that existing detections may represent only the tip of the underlying population.

    \item \textbf{SMBH population and emergence of quenched galaxies}: SMBHs at intermediate masses ($M_{\rm BH}\lesssim 10^{8}\msun$) are more abundant in EDE, whereas the most massive ones are less massive owing to the shorter cosmic age in EDE and the rapid SMBH growth timescale at the Eddington limit. Since the onset of quenching depends critically on the most massive SMBHs and their accumulated kinetic-mode feedback energy (in the IllustrisTNG model), SMBHs cross the critical mass for quenching galaxies at later times in EDE, leaving a later emergence of quenched galaxies at roughly $z\simeq 3.5$ as opposed to $z\simeq 4$ in $\Lambda$CDM. The resulting number densities remain below observational estimates, highlighting a limitation of our SMBH model.
\end{itemize}

These results demonstrate that EDE provides a physically motivated cosmological framework that naturally reproduces multiple statistical properties of high-redshift galaxy populations, including UV luminosity functions, stellar mass functions, massive galaxy outliers, and disk fractions. No fine-tuning is required for either the adopted galaxy formation model or the cosmological model constrained independently by CMB. The empirical success of IllustrisTNG at late times is preserved as the EDE predictions converge to $\Lambda$CDM at lower redshifts. This strengthens the case for EDE as a viable extension to the standard $\Lambda$CDM model, while the remaining challenge of early quenching likely indicates limitations of the subgrid SMBH physics. From a broader perspective, our findings highlight how modifications to the pre-recombination expansion history can leave measurable imprints on high-redshift galaxy formation. EDE serves here as a representative example.

Significant degeneracies remain among various baryonic and ``dark'' physics solutions proposed to address early-universe anomalies. A promising path forward is to place joint constraints on these uncertain physical processes through galaxy clustering measurements. In the case of EDE, clustering is expected to reflect several interrelated effects: an enhancement in the underlying dark matter correlation function, a reduction in halo bias (since halos of fixed mass correspond to less extreme, more abundant perturbations in the early universe), and modifications to the galaxy-halo connection required to match the observed UV luminosity and stellar mass functions. Some of these observable consequences have already been illustrated in Figure~\ref{fig:img}. To fully capture these effects and translate them into quantitative constraints, larger-volume simulations that explore a wider parameter space of the galaxy formation model will be necessary. Additionally, incorporating on-the-fly radiative transfer into such simulations will enable studies of the topology of neutral/ionized gas during the epoch of reionization in alternative cosmologies. Opportunities remain to leverage Lyman-$\alpha$ forest and 21cm observations to constrain the small-scale matter power spectrum. While these probes have been extensively used to rule out models that suppress small-scale power, they have yet to be fully explored in the opposite regime, where the power spectrum is enhanced as in the EDE scenario.

\section*{Acknowledgements}
Simulations in this work are conducted on the Cannon cluster managed by the Faculty of Arts and Sciences Research Computing at Harvard University. Post-processing is done on the Engaging cluster at Massachusetts Institute of Technology. XS acknowledges the support from the National Aeronautics and Space Administration (NASA) theory grant JWST-AR-04814.
Support for OZ was provided by Harvard University through the Institute for Theory and Computation Fellowship. MBK acknowledges the support from the National Science Foundation (NSF) grants AST-1910346, AST-2108962, and AST-2408247; NASA grant 80NSSC22K0827; HST-GO-16686, HST-AR-17028, HST-AR-17043, JWST-GO-03788, and JWST-AR-06278 from the Space Telescope Science Institute, which is operated by AURA, Inc., under NASA contract NAS5-26555; and from the Samuel T. and Fern Yanagisawa Regents Professorship in Astronomy at UT Austin.

\section*{Data Availability}
The simulation and post-processing data can be shared upon request to the corresponding author of this paper.





\appendix

\section{Halo mass function}

\begin{figure}
    \centering
    \includegraphics[width=\linewidth]{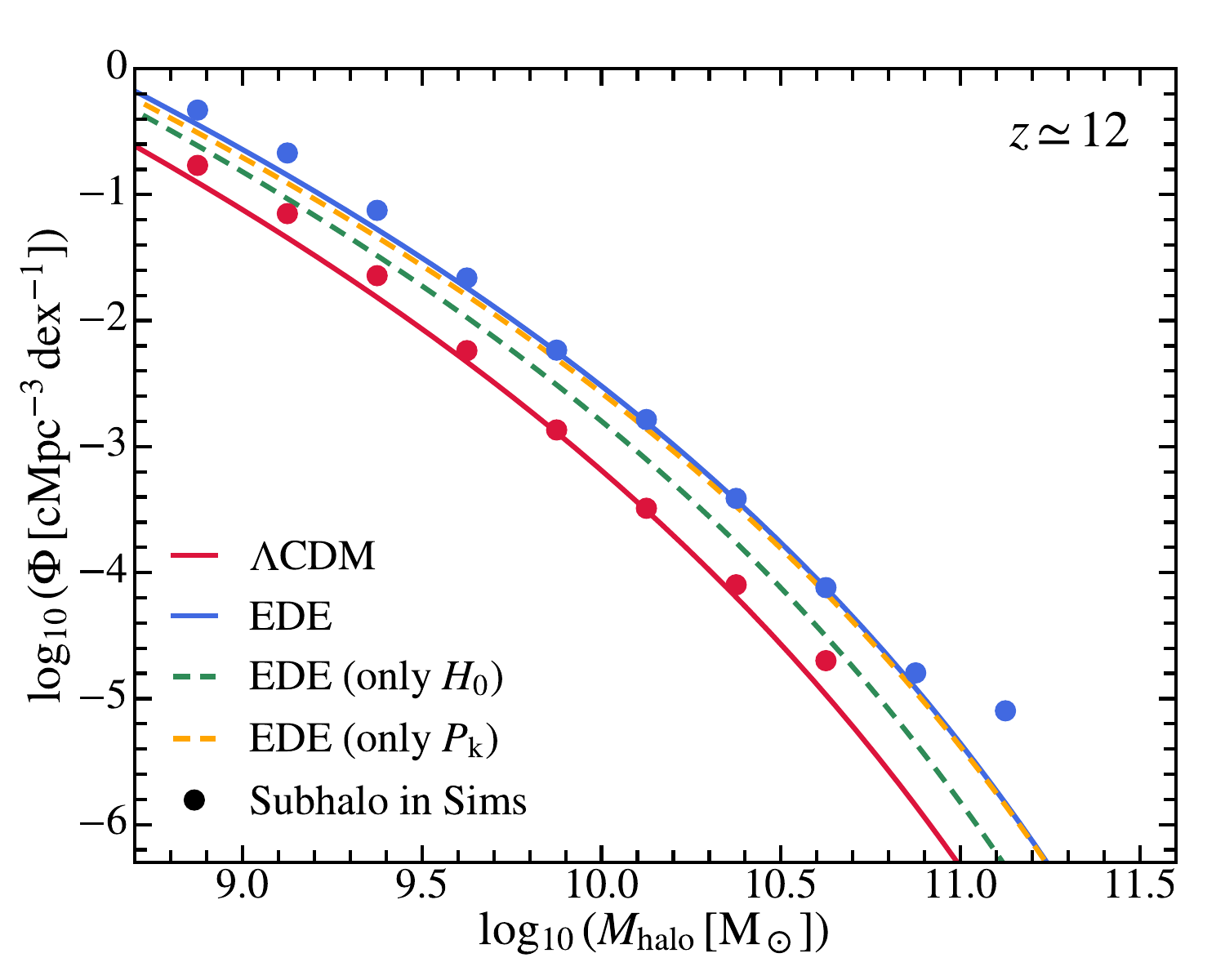}
    \caption{Halo mass function at $z\simeq 12$. Solid lines show predictions from the empirical model for both the $\Lambda$CDM and EDE cosmologies. The green dashed line isolates the effect of changing only $H_0$ to the EDE value, while the orange dashed line reflects the impact of replacing only the linear matter power spectrum with that of EDE. Data points represent binned estimations of the subhalo mass function from our simulations. The empirical model closely reproduces the simulation results. The enhanced halo abundance in EDE is primarily driven by changes in the linear matter power spectrum as found here.
    }
    \label{fig:hmf_z12}
\end{figure}

Figure~\ref{fig:hmf_z12} presents the halo mass function at $z\simeq 12$ predicted by our empirical model and measured directly from simulations. The model predictions for both the $\Lambda$CDM and EDE cosmologies are shown as solid lines, and are in excellent agreement with the mass functions of subhaloes identified from the simulations using \textsc{Subfind-hbt}. To disentangle the physical drivers of the enhanced halo abundance in EDE, we show two intermediate cases: one where only the Hubble constant is modified to match the EDE value (green dashed line), and another where only the linear matter power spectrum is replaced with that of the EDE model (orange dashed line), keeping all other parameters fixed. The comparison demonstrates that the dominant contribution to the difference in halo abundance between EDE and $\Lambda$CDM arises from changes in the linear matter power spectrum. These results also highlight the effectiveness of our empirical model in capturing early structure formation across different cosmologies.

\section{Disk mass fraction}

\begin{figure}
    \centering
    \includegraphics[width=\linewidth]{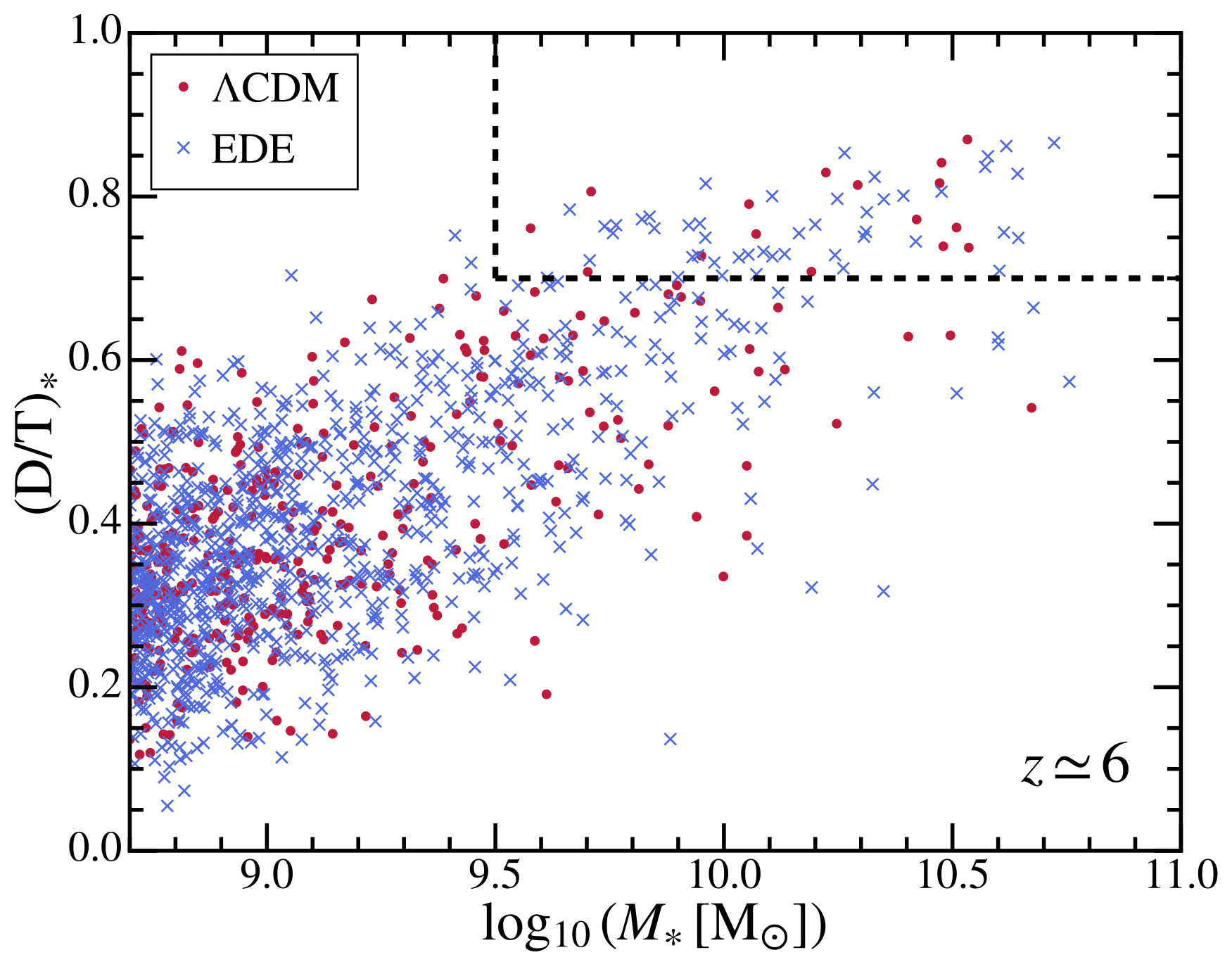}
    \includegraphics[width=\linewidth]{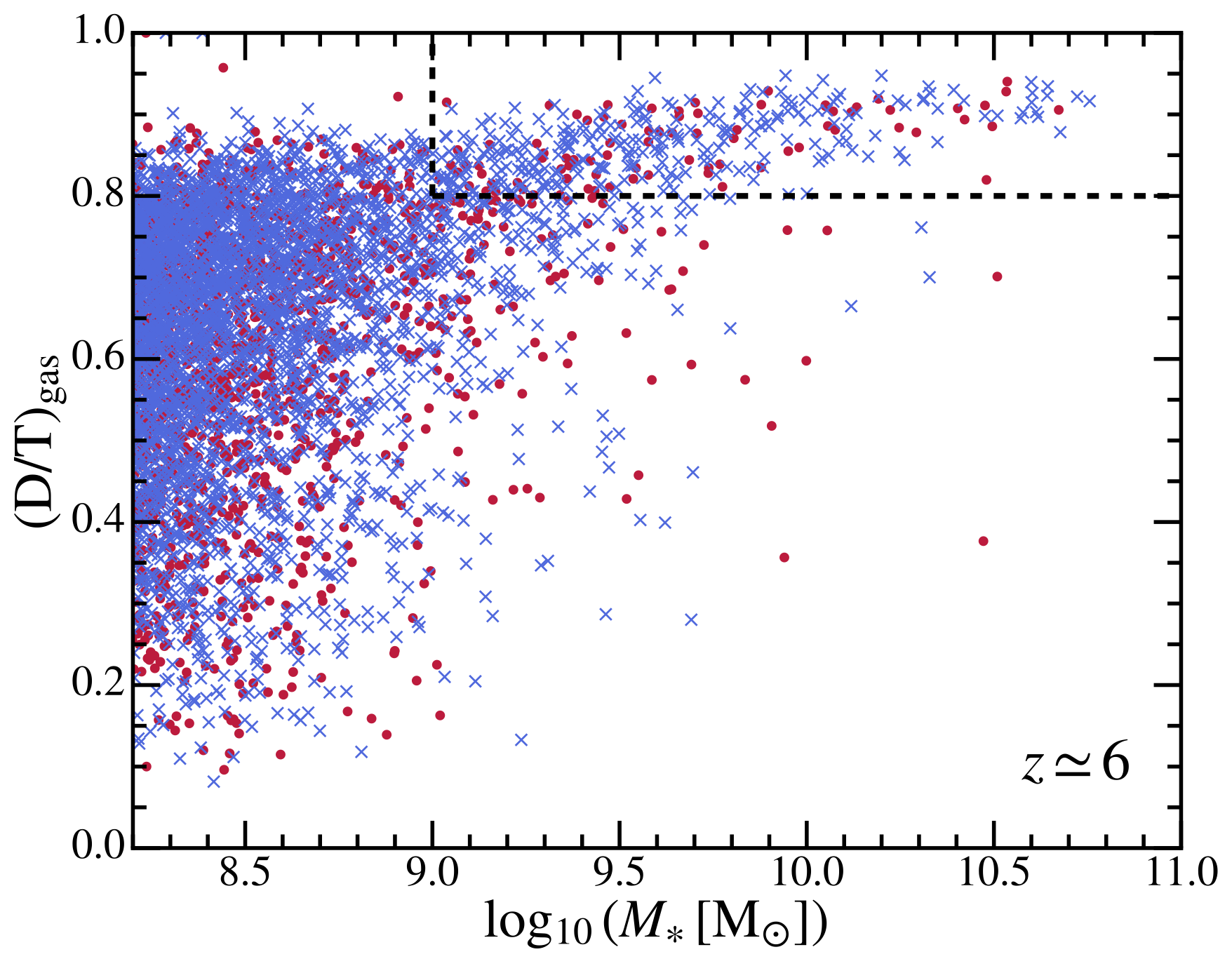}
    \caption{Fraction of stellar (gas) mass identified in the disk component as a function of galaxy stellar mass at $z\simeq 6$, shown in the top (bottom) panel. We overplay the criteria where a galaxy is considered to have a stellar or gaseous disk in dashed lines.}
    \label{fig:fdisk_z6}
\end{figure}

In Figure~\ref{fig:fdisk_z6}, we show the disk-to-total mass ratio (D/T) as a function of stellar mass for galaxies in our $\Lambda$CDM and EDE simulations at $z\simeq 6$. We show the distribution of galaxies relative to the thresholds we adopt for identifying ``disky'' galaxies. The stellar D/T exhibits a strong positive correlation with galaxy stellar mass. As discussed in the main text, our fiducial threshold of (D/T)$\ast = 0.7$ is motivated by kinematic studies of Milky Way–like galaxies at lower redshift in simulations employing the same galaxy formation model. With this definition, the classification is largely insensitive to the stellar mass cut we apply, which was originally chosen to ensure consistency with the mass completeness of the observational samples. The gas component generally exhibits higher D/T values. We adopt a threshold of (D/T)${\rm gas} = 0.8$, selected to include all prominent gas disks at the massive end. The stellar mass cut shown in the figure is somewhat arbitrary, chosen primarily to match the $10^{9}-10^{10} \msun$ range of galaxies observed with dynamically cold gas disks by ALMA.

\begin{figure*}
    \raggedright
    \includegraphics[width=0.33\linewidth]{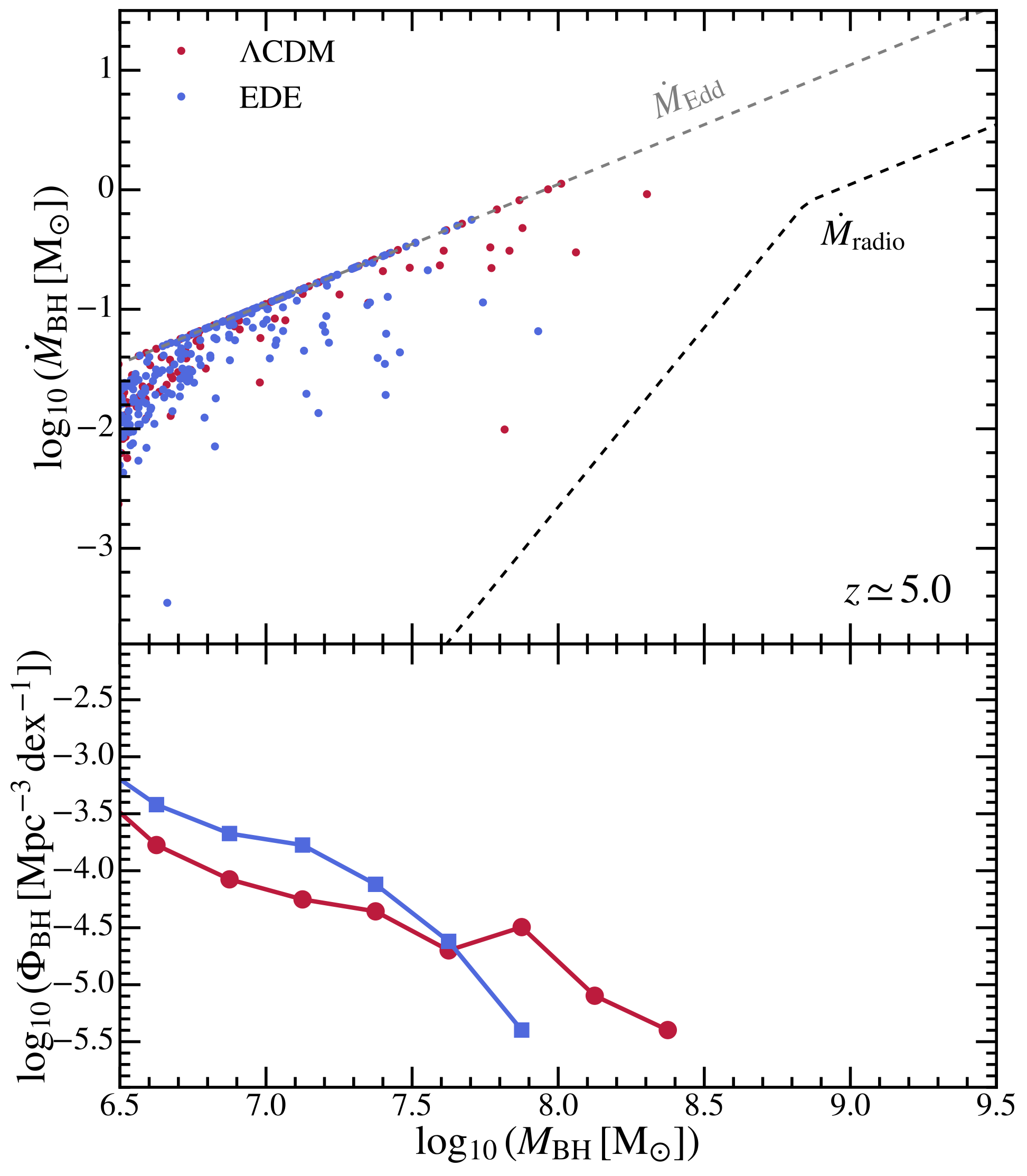}
    \includegraphics[width=0.33\linewidth]{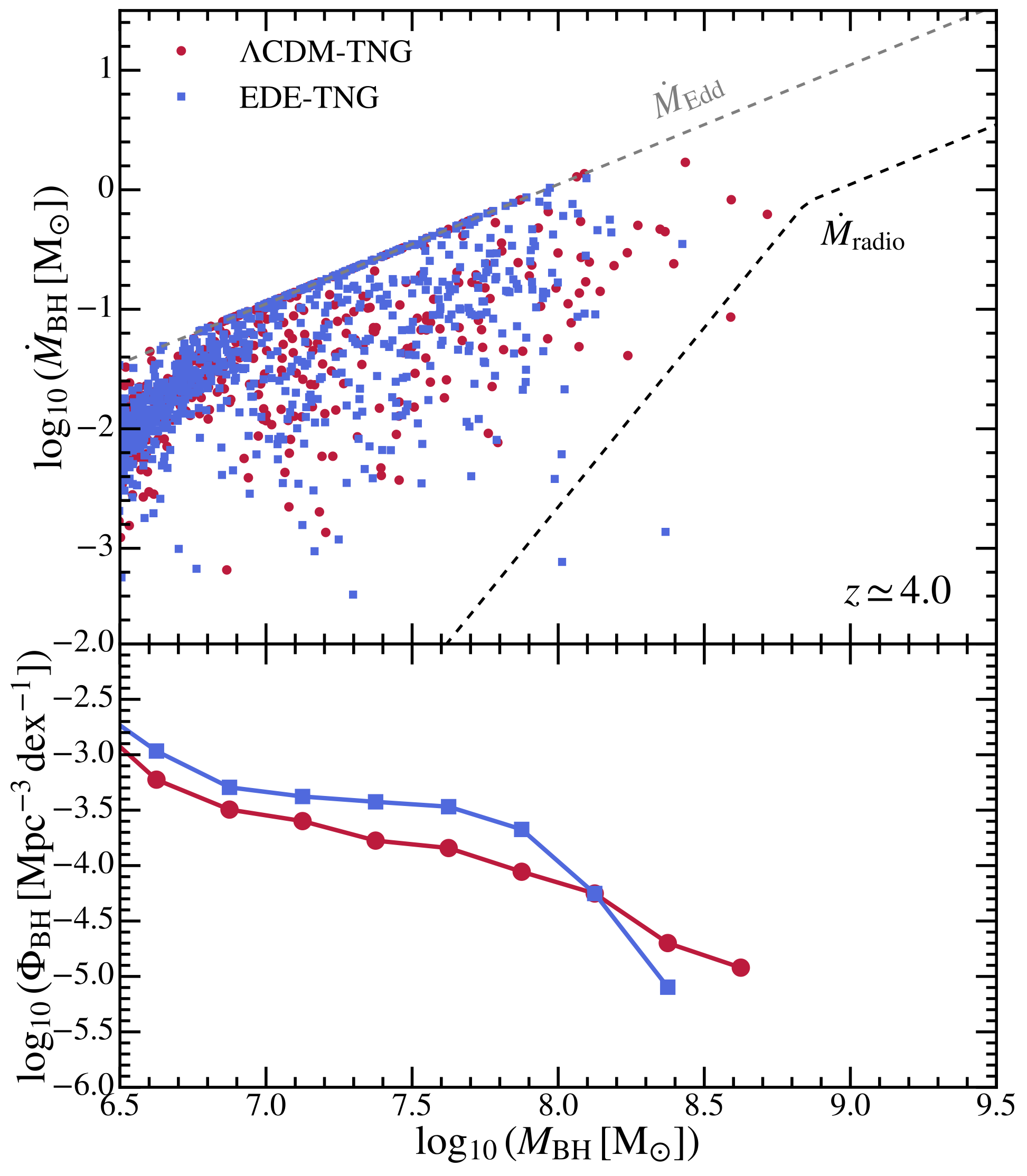}
    \includegraphics[width=0.33\linewidth]{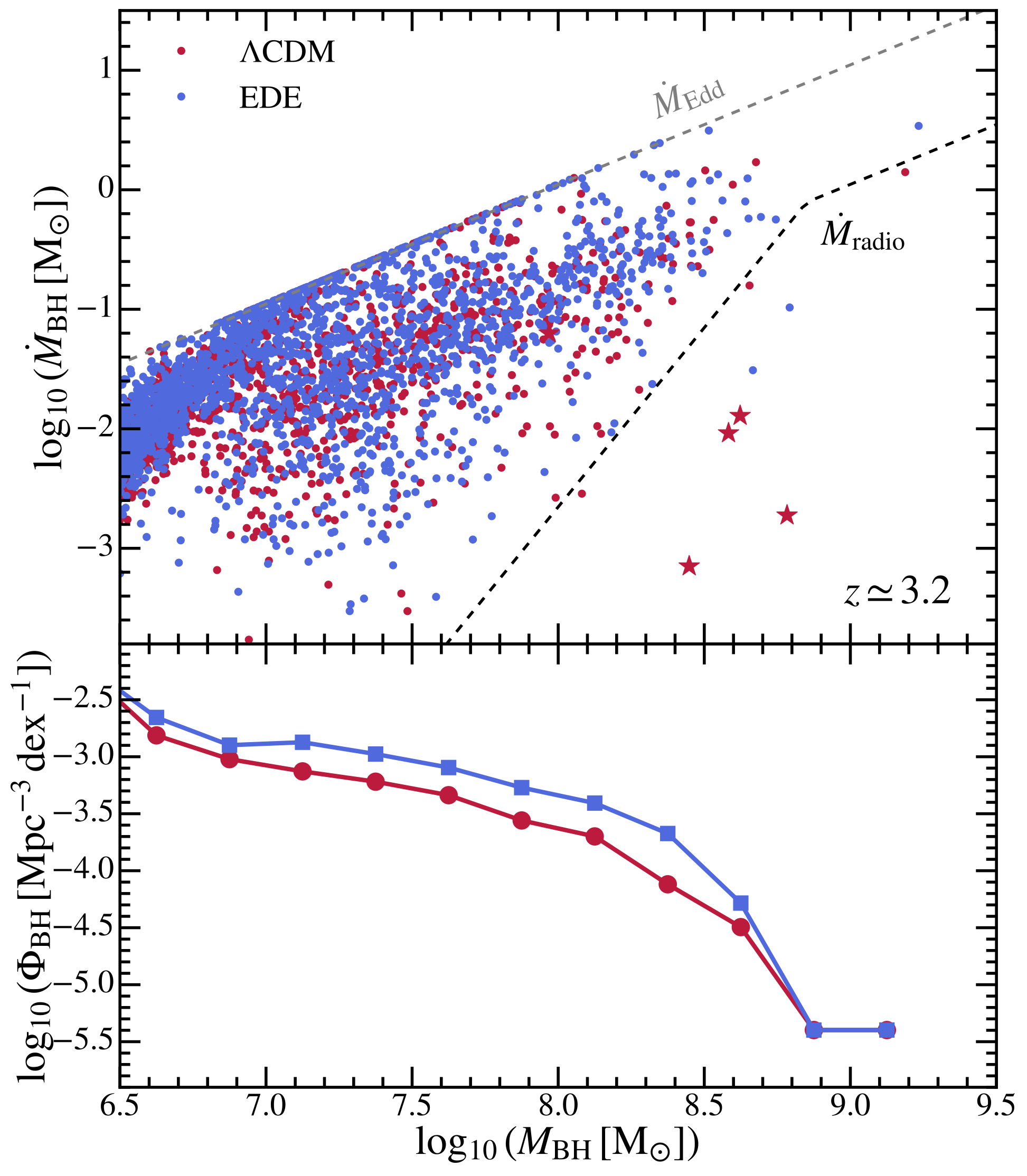}
    \includegraphics[width=0.33\linewidth]{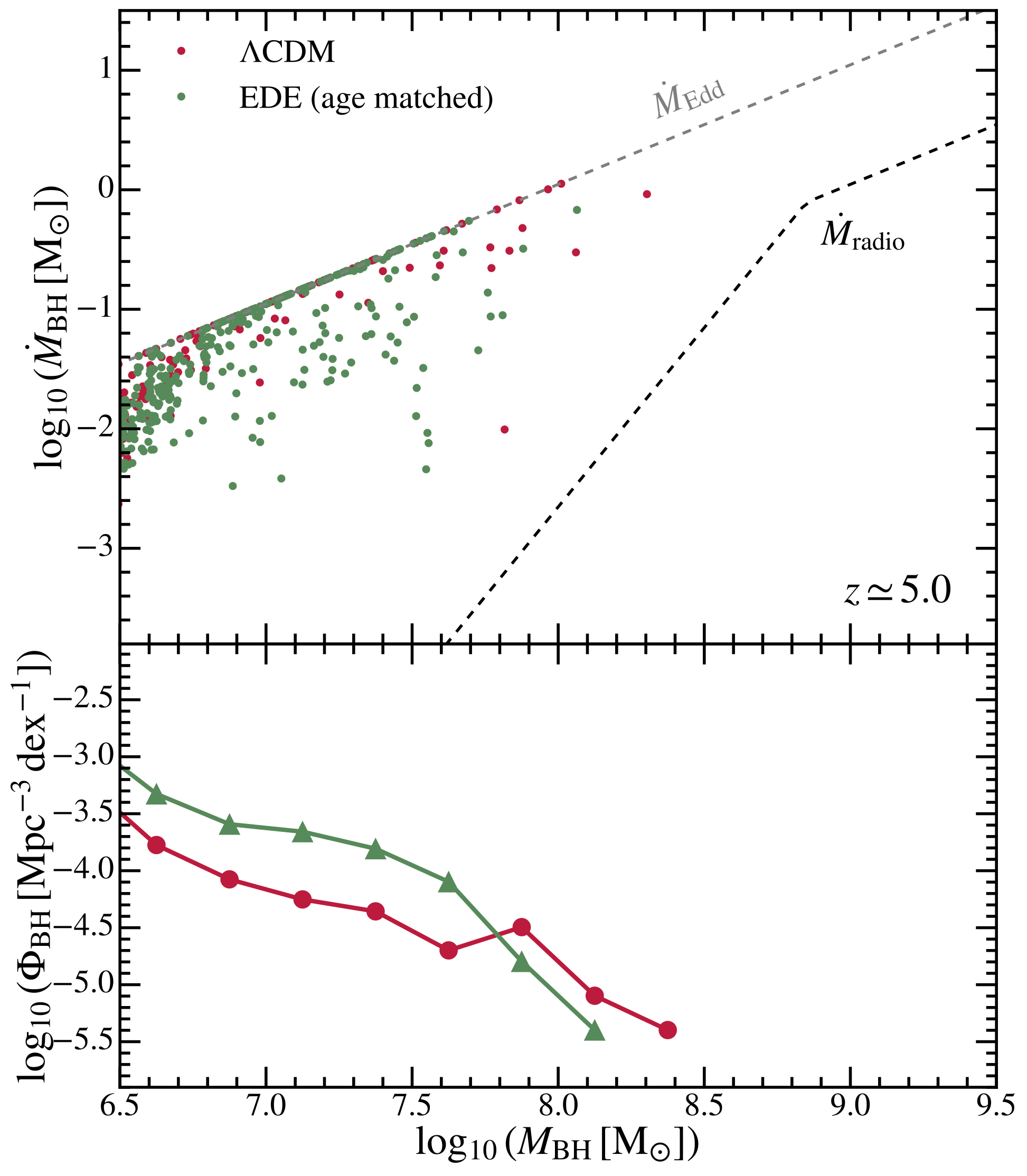}
    \includegraphics[width=0.33\linewidth]{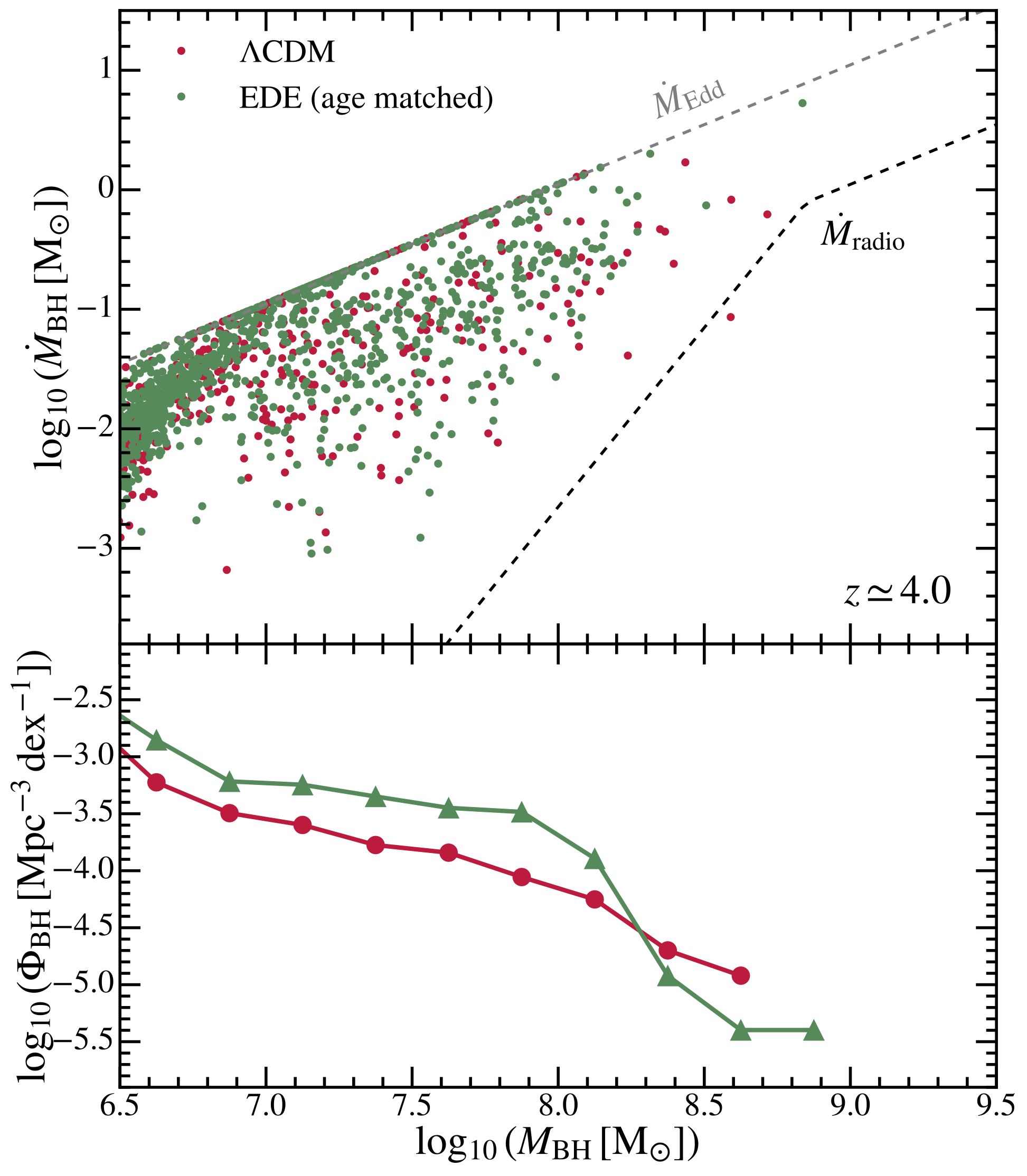}
    \includegraphics[width=0.33\linewidth]{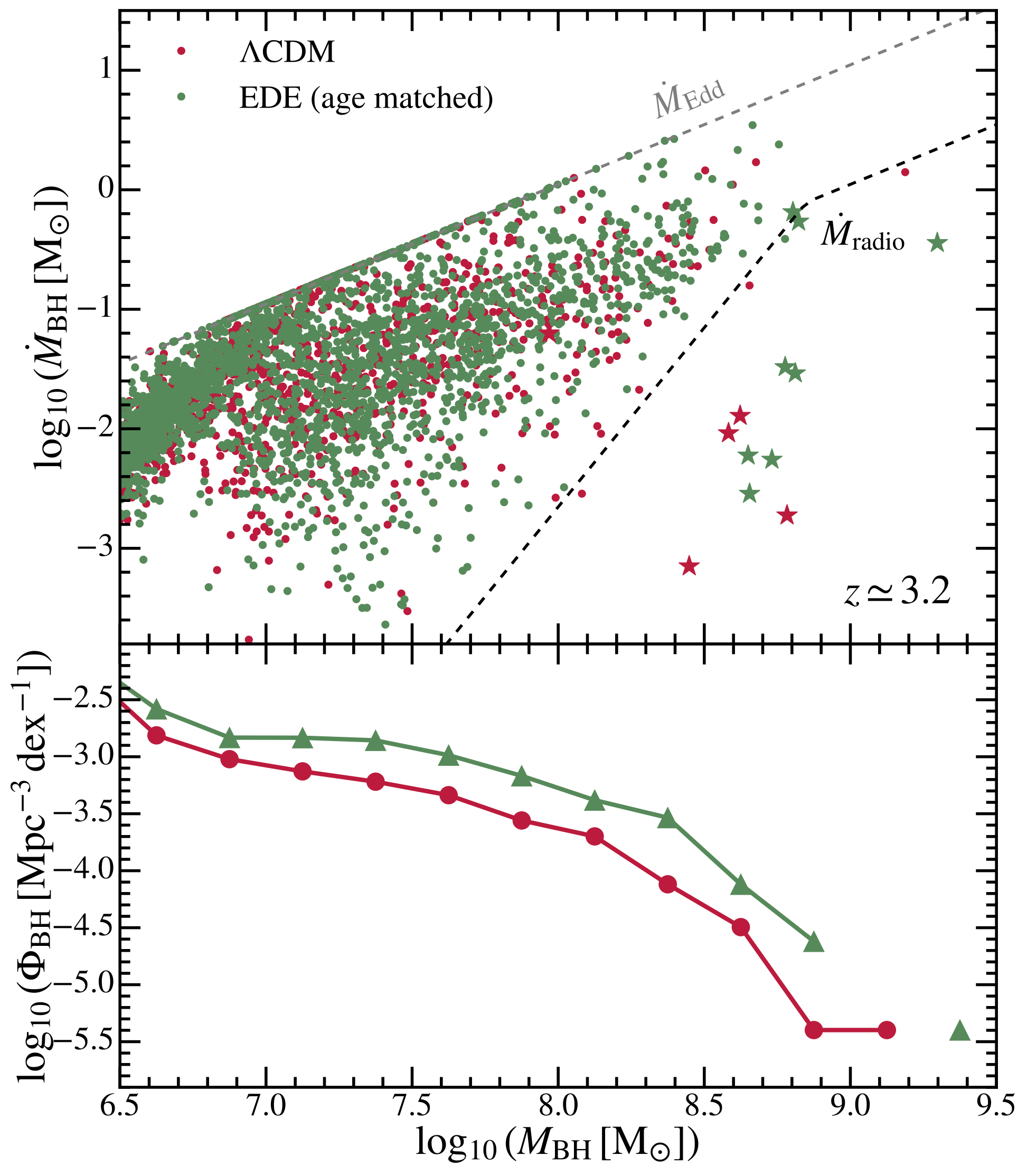}
    \caption{In each plot, the top subpanel shows the SMBH accretion rate versus mass, and the bottom subpanel shows the SMBH mass function. We perform two sets of comparisons. The first row of plots compares $\Lambda$CDM and EDE predictions at the same redshift. The second row of plots compares them at the same age of the universe, and is denoted as ``age matched''. We overlay the Eddington-limit accretion rates and the limit where the kinetic mode AGN feedback is triggered. Galaxies marked by stars are quenched galaxies.}
    \label{fig:mdot_and_phibh}
\end{figure*}

\section{The evolution of SMBHs}

In Figure~\ref{fig:mdot_and_phibh}, we present the SMBH mass accretion rate as a function of SMBH mass in our $\Lambda$CDM and EDE simulations at $z\simeq 5$, 4, and 3.2. The bottom subpanels display the corresponding SMBH mass functions. At $z\simeq 5$, SMBHs are accreting near the Eddington limit. As they grow, their accretion rates decline to more moderate levels, typically around 10\% of the Eddington rate, as regulated by thermal-mode AGN feedback. A sharp drop in accretion occurs when SMBHs transition to the kinetic feedback mode, triggering the quenching of their host galaxies soon after. This transition typically occurs at a characteristic SMBH mass of $M_{\rm BH} \simeq 10^{8.5} \msun$.

Due to the rapid early growth of SMBHs, their final mass is highly sensitive to the available time for accretion since seeding, which in turn depends on the age of the universe in a given cosmology. The EDE model results in a universe that is $50$ to $100 \Myr$ younger at $z\gtrsim 3$, reducing the time available for SMBH growth. As a result, the abundance of the most massive SMBHs (e.g. $M_{\rm BH}>10^{8}\msun$) is slightly lower in EDE, despite a generally higher abundance at low and intermediate masses. This shift has a direct impact on the timing of kinetic-mode feedback activation and, consequently, galaxy quenching.

To further isolate this effect, the bottom row of Figure~\ref{fig:mdot_and_phibh} compares $\Lambda$CDM and EDE results at matched cosmic age rather than redshift. In this frame, the massive end of the SMBH mass function in EDE becomes more consistent with that of $\Lambda$CDM. By $z \simeq 3.2$, both cosmologies produce a comparable population of quenched galaxies, and the EDE model catches up with or even surpasses $\Lambda$CDM in terms of massive SMBH abundance at fixed age.


\bsp	
\label{lastpage}
\end{document}